\documentclass[times,twocolumn,final,authoryear]{elsarticle}

\usepackage{jasr}
\usepackage{framed,multirow}
\usepackage{amssymb}
\usepackage{latexsym}
\usepackage{url}
\usepackage{xcolor}
\usepackage[citebordercolor=white]{hyperref}

\journal{Advances in Space Research}
\graphicspath{{figures/}}

\definecolor{darkgreen}{rgb}{0., 0.5, 0.4} %
\definecolor{pyellow}{rgb}{1, 0.8, 0.} %

\newcommand{\p}[1]{{\color{magenta}{#1}}}

\newcommand{\BE}{\begin{equation}}
\newcommand{\EE}{\end{equation}}
\newcommand{\BA}{\begin{align}}
\newcommand{\EA}{\end{align}}
 \newcommand{\fig}[1]{Figure~\ref{fig_#1}}

 \newcommand{\figs}[2]{Figures~\ref{fig_#1} and \ref{fig_#2}}
 
 \newcommand{\sect}[1]{Section~\ref{sect_#1}}

 \newcommand{\tabl}[1]{Table~\ref{tabl_#1}}

\newcommand{\degree}{\ensuremath{^\circ}}

\newcommand{\insitu}{{\it in situ}}

\newcommand{\HCME}{H_{\rm CME}}
\newcommand{\tauCME}{\tau_{\rm CME}}
\newcommand{\phitCME}{\phi_{\rm t,CME}}
\newcommand{\Pmag}{P_{\rm mag}}
\newcommand{\Pth}{P_{\rm th}}

\newcommand{\obs}{$_{\rm obs}$}
\newcommand{\simu}{$_{\rm simu}$}
\newcommand{\simuone}{ $_{\rm simu,1}$}
\newcommand{\simutwo}{$_{\rm simu,2}$}

\begin{document}
\verso{Verbeke \textit{etal}}

\begin{frontmatter}

\title{
Over-expansion of coronal mass ejections modelled using 3D MHD EUHFORIA simulations}

\author[1,2]{Christine \snm{Verbeke} } 
\author[2,3]{Brigitte \snm{Schmieder}} 
\author[3,4]{Pascal \snm{D\'emoulin}} 
\author[5,6,7]{Sergio \snm{Dasso}}
\author[8]{Benjamin \snm{Grison}}
\author[1,2]{Evangelia \snm{Samara}}
\author[9,10]{Camilla \snm{Scolini}}
\author[2,11]{Stefaan \snm{Poedts}}
\address[1]{Royal Observatory of Belgium, Ringlaan 3, 1180 Ukkel, Belgium, cgjmverbeke@gmail.com}
\address[2]{CmPA/Department of Mathematics, KU Leuven, Celestijnenlaan 200 B, 3001 Leuven, Belgium}
\address[3]{LESIA, Observatoire de Paris, Universit\'e PSL, CNRS, Sorbonne Universit\'e, Univ. Paris Diderot, Sorbonne Paris Cit\'e, 5 place Jules Janssen, 92195 Meudon, France}
\address[4]{Laboratoire Cogitamus, rue Descartes, 75005 Paris, France}
\address[5]{Universidad de Buenos Aires, Facultad de Ciencias Exactas y Naturales, Departamento de Ciencias de la Atm\'osfera y los Oc\'eanos, 1428 Buenos Aires, Argentina}
\address[6]{CONICET, Universidad de Buenos Aires, Instituto de Astronom\'\i a y F\'\i sica del Espacio, CC. 67, Suc. 28, 1428 Buenos Aires, Argentina}
\address[7]{Universidad de Buenos Aires, Facultad de Ciencias Exactas y Naturales, Departamento de Física, 1428 Buenos Aires, Argentina}
\address[8]{Department of Space Physics, Institute of Atmospheric Physics of the Czech Academy of Sciences, Prague, Czech Republic}
\address[9]{Institute for the Study of Earth, Oceans, and Space, University of New Hampshire, Durham, NH 03824, USA}
\address[10]{CPAESS, University Corporation for Atmospheric Research, Boulder, CO 80301, USA}
\address[11]{Institute of Physics, University of Maria Curie-Sk{\l}odowska, Pl. M. Curie-Sk{\l}odowskiej 5, 20-031 Lublin, Poland}

\received{1 October 2021}
\finalform{2021}
\accepted{2021}
\availableonline{2021}

\begin{abstract}
\textit{Context:} Coronal mass ejections (CMEs) are large scale  eruptions observed close to the Sun. They are travelling through the heliosphere and possibly interacting with the Earth environment creating interruptions or even damaging new technology instruments. Most of the time their physical conditions (velocity, density, pressure) are only measured \insitu\ at one point in space, with no possibility to have information on the variation of these parameters during their journey from Sun to Earth. \\
\textit{Aim:} Our aim is to understand the evolution of internal  physical parameters of a set of three particular fast halo CMEs. These CMEs were launched between 15 and 18 July 2002.  
Surprisingly, the related interplanetary CMEs (ICMEs), observed near Earth, have a low, and in one case even very low, plasma density.  \\
\textit{Method:} We use the EUropean Heliosphere FORecasting Information Asset (EUHFORIA) model to simulate the propagation of the CMEs in the background solar wind by placing virtual spacecraft along the Sun--Earth line. We set up the initial conditions at 0.1 au, first with a cone model and then with a linear force free spheromak model.  \\ 
\textit{Results:} 
A relatively good agreement between simulation results and observations concerning the speed, density and arrival times of the ICMEs is obtained by adapting the initial CME parameters. In particular, this is achieved by increasing the initial magnetic pressure so that a fast expansion is induced in the inner heliosphere. This implied the development of fast expansion for two of the three ICMEs.  In contrast, the intermediate ICME is strongly overtaken by the last ICME, so that its expansion is strongly limited.\\
\textit{Conclusions:}
      First, we show that a magnetic configuration with an out of force balance close to the Sun mitigates the EUHFORIA assumptions related to an initial uniform velocity.  {Second, the over-expansion of the ejected magnetic configuration in the inner heliosphere is one plausible origin for the low density observed in some ICMEs at 1 au. Furthermore, we conclude for one ICME, surrounded by two other ICMEs, that the insitu observed very low density has a possible coronal origin.}


\end{abstract}


\end{frontmatter}

\section{Introduction} 
\label{sect_Introduction}
%
Coronal mass ejections (CMEs) are important drivers of space-weather disturbances observed at Earth \citep{illing1985,webb2012}.  CMEs are initiated in the low solar corona by the launch of eruptive solar plasma and magnetic fields. Those that propagate with a speed higher than the speed of the ambient solar wind may drive a shock ahead of them. The compressed plasma accumulated in front of the magnetic ejecta (ME) is called the sheath of the CME. After travelling through the heliosphere as interplanetary CMEs (ICMEs), they may arrive at Earth and as such they meet the spacecraft at the Sun-Earth Lagrangian point L1 where the \insitu\ speed, density and magnetic field of the solar wind and embedded ICMEs can be measured.
Applying triangulation techniques from coronagraph images, the shape of CMEs can be retrieved \citep[see e.g.][] {Mierla2008, Braga2017, Balmaceda2018}. Using the estimated CME kinematics, we can model and study the ICME arrival time at Earth using a wide variety of ICME propagation models. \\

Up until now it has been difficult to find propagating CMEs that are observed \insitu\ by multiple spacecraft \citep[see, e.g., the recent inner heliospheric catalogs by][]{Grison18,Good2019, Salman2020}, in order to study the evolution of their shape and structure into more detail. Recently, a multi-spacecraft encounter has been studied by also simulating the event using magnetohydrodynamical (MHD) simulations  \citep{Asvestari2021}. Even after the launch in 2018 of NASA’s Parker Solar Probe \citep[PSP,][]{Fox2016}, such events are rare \citep{Davies2020, Winslow2021, Moestl2022}, as we are currently in a phase of low solar activity.
Furthermore, studies have shown that the CME expansion in the inner heliosphere is driven mainly by the decrease of solar wind pressure with distance \citep{Demoulin2009,Gulisano2010}. Finally, \citet{Lugaz2020} have found that the expansion rate depends on the initial magnetic field strength which is an indirect evidence that CME expansion close to the Sun is driven by the internal magnetic pressure.\\

Previous studies indicated that several spacecraft have already observed  periods of very low density in the solar wind \citep{Lugaz2016,Chane2021,Rajkumar2022}.
\citet{Chane2021} provided an explanation of such a low density observed between May 24-25 2002 by over-expansion
of an ICME during its journey to Earth. Because the ICME was travelling in the wake of another CME ejected just a few hours earlier, i.e., the preceding ICME cleared away the ambient solar wind
plasma and the frozen-in magnetic field, the observed over-expansion was made possible.
%
%
In the present paper, we study two other cases of low density ICMEs observed on 18-20 July 2002.
They belong to the twelve fast halo CMEs  associated with twelve X-ray flares observed in 2002. These twelve CMEs have been studied previously in order to understand why they were not geo-effective \citep{Schmieder2020}. It was found that their sources were mainly in large sunspot groups but per chance located close to or at the west or east limb and that the orientation of the interplanetary magnetic field (IMF) was mainly northward at L1, so not geo-effective.
%
To complement the observations, we have performed numerical simulations using the EUHFORIA model \citep[EUropean Heliosphere FORecasting Information Asset][]{Pomoell2018}, which allow us to track the ICMEs as they travel through the inner heliosphere. We follow the techniques described by \citet{Scolini2021} to track the background solar wind and the ICMEs traveling through it by using virtual spacecraft.
In order to initiate the simulations we use two different CME models: the cone model which represents a hydrodynamic pulse and the Linear Force Free (LFF) spheromak model \citep{Verbeke2019b}, which includes an intrinsic magnetic field.
Furthermore, we derived the expansion rate from the linear profile of the observed velocity profile, assuming a nearly self-similar expansion through the heliosphere, similar as in \citet{Chane2021}.\\

The present paper is organised as follows. In \sect{Observations}, we present the solar observations between 15-18 July 2002, then the \insitu\ measurements at L1 are analysed.  We describe the simulation set-up and results in \sect{simulations} and we analyse the simulated ICME between 0.1 and 1 au in \sect{evolution}. Finally, discussion and conclusions are presented in \sect{Discussion}.

\section{Observations : CME sources and ICMEs at L1}
\label{sect_Observations}

\begin{figure*}[ht!]
\centering
\includegraphics[scale=0.8]{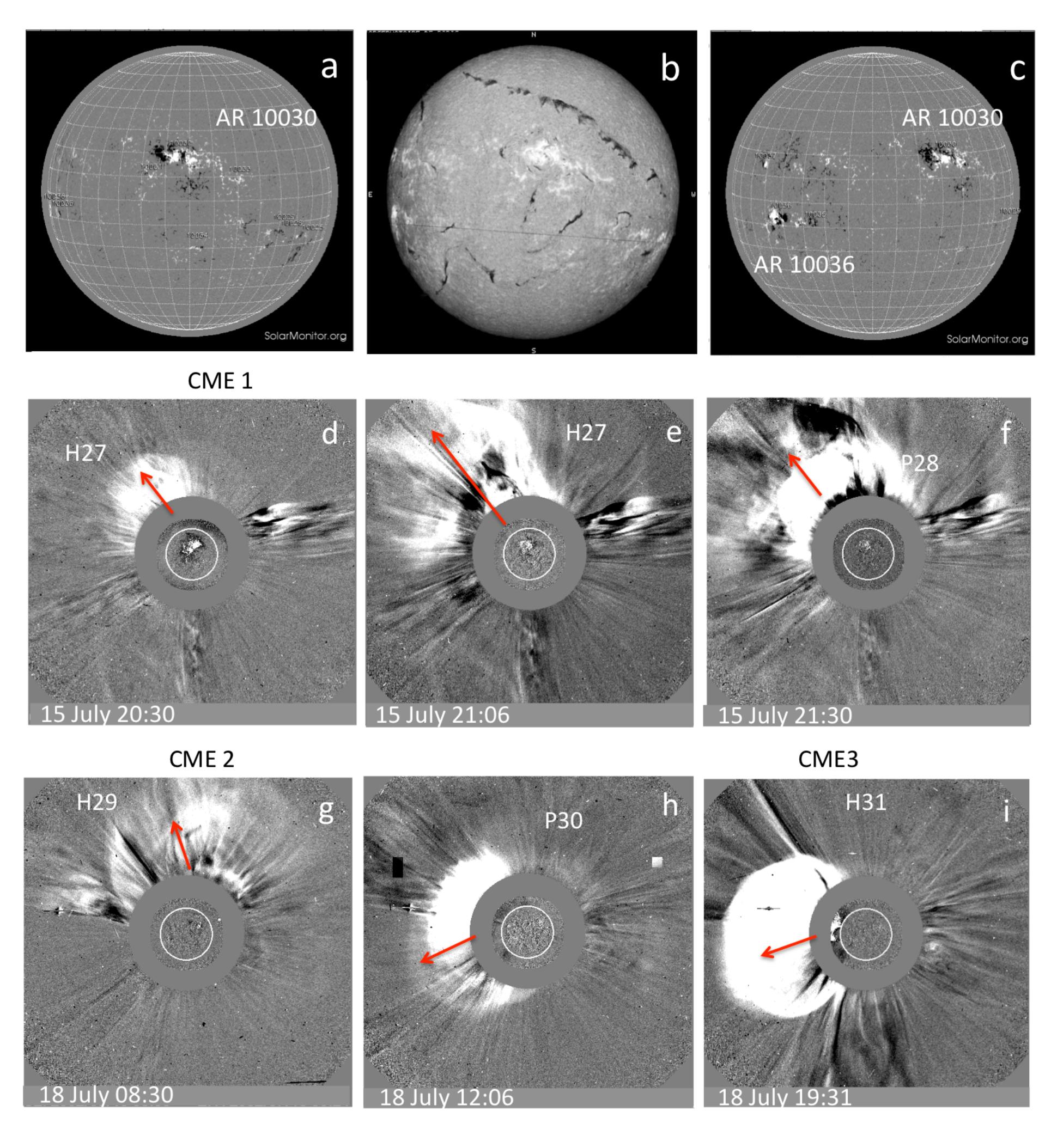}
\caption{
Observations of the solar sources of the CMEs in AR 10030 and AR 10036 between 15 and 18 July 2002. (a, c) Data of HMI at 12:51~UT on 15 July 2002 and 12:47~UT on 18 July 2002, respectively.  (b)  Meudon spectroheliogram in H$\alpha$ at 9:30 UT on 16 July 2002. 
(d, e, f, g, h, i) Halo  and partial CMEs observed by LASCO C2 on 15 and 18 July 2002. The red arrows indicate the projected direction of the CMEs.
}
\label{fig_radio}
\end{figure*}

\begin{figure*}[ht!]
\centering
\includegraphics[scale=0.9]{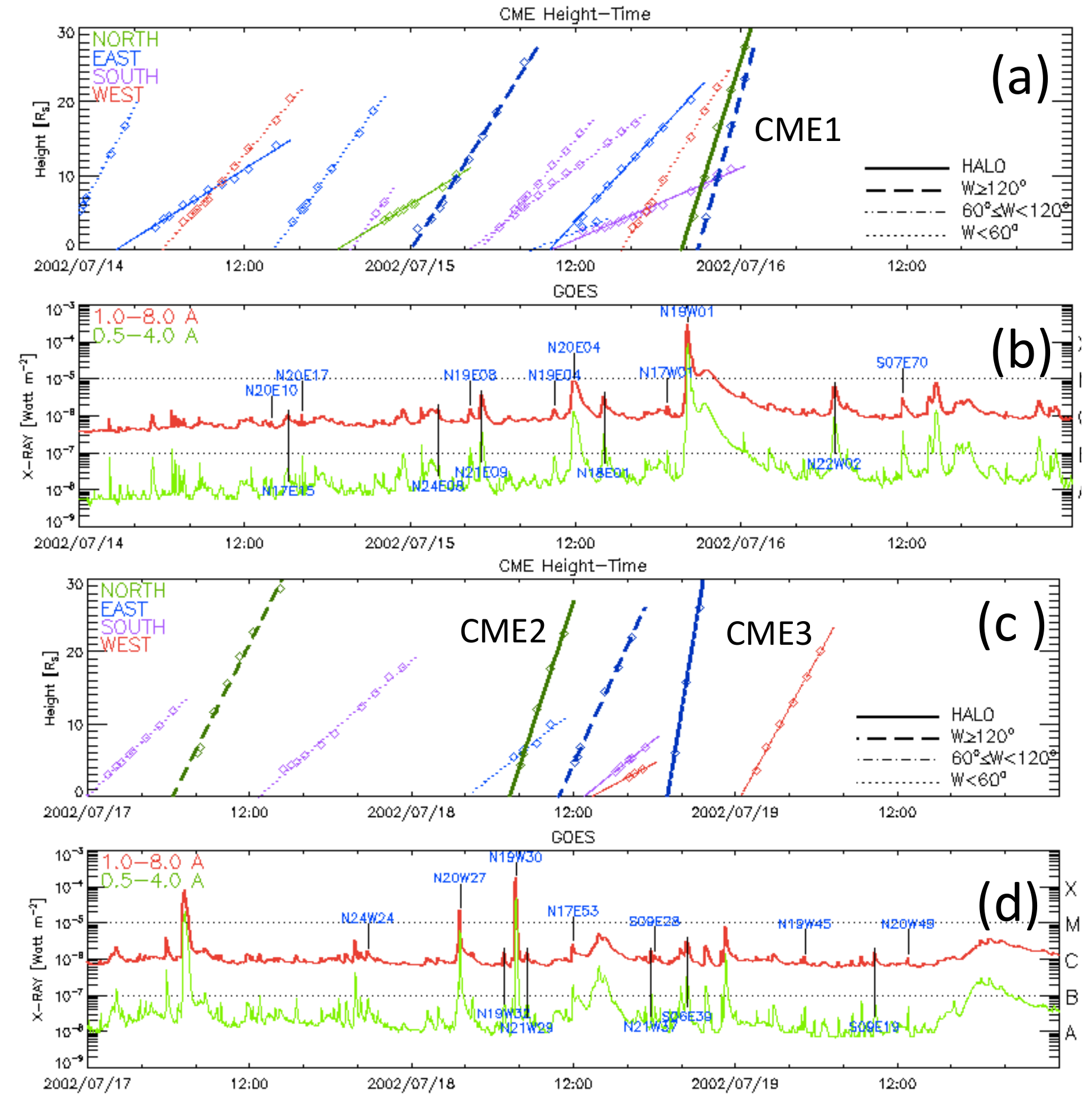}
\caption{CMEs characteristics. 
(a,c) Height time plots where the three CMEs studied in this paper are marked with continuous thick lines. The colors codes the main CME direction (see left insert), while the type of lines codes the CME apparent angular extension (see right insert). 
(b,d) GOES X-rays flux in two channels with the solar source coordinates of the identified events. The two top panels are for the time range 14-16 July 2002, and the bottom panels for 17-19 July 2002.  }
\label{fig_CME}
\end{figure*}

\subsection{Sources of the CMEs in July 2002}
\label{sect_Obs_Sources}

Over the span of a few days between 15 and 18 July 2002, multiple CMEs were observed.
For the purpose of this study, we focus on a set of three CMEs whose ICME counterparts have already been studied in detail in two successive papers \citep{Bocchialini2018,Schmieder2020}. A summary of their properties is provided in Table~\ref{tabl_cme_lasco}.
\begin{table*}[htb!]
\centering
\begin{tabular}{llllllrl}
\hline
  CME   & CME \#    & Day   & Time &  Heliographic   & AR   & Speed         & Flare \\
     &           &       & [UT] & coordinates    & NOAA & [km s$^{-1}$] & class\\
\hline
H27 & CME1  & 15 July    & 19:59 & (E04, N15) & 10030 & 974  & X\\
P28 &       & 15 July    & 21:00 & (W01, N15) & 10030 & 1274 & M\\
H29 & CME2  & 18 July    & 07:59 & (W30, N15) & 10030 & 919  & X\\
P30 &       & 18 July    & 11:30 & (E10, S40) & 10036 & 680  & C\\
H31 &CME3   & 18 July    & 18:26 &(E10, S40)  & 10036 & 1788 & C\\
 \hline
\end{tabular}
\caption{CMEs observed between 15 and 18 July 2002 by LASCO C2 at 6~$R_\odot$ and their solar sources and related flares in heliographic coordinates (North, South, East, West, +/- 90 degrees). 
Time and speed information is taken from the CDAW CME catalogue. CME1, CME2, CME3 are halo CMEs related to a sudden storm commencement at the Earth \citep{Bocchialini2018}. 
In the first column we recap the CME name given by \citet{Bocchialini2018} for reference.}
\label{tabl_cme_lasco}
\end{table*}
We present here data from the Michelson Doppler Imager (MDI) instrument (see \url{https://www.solarmonitor.org/}) as well as coronagraph images from the Large Angle and Spectrometric Coronagraph \citep[LASCO,][]{Brueckner1995} experiment providing images of the inner solar corona. These instruments are both on board of the Solar and Heliospheric Observatory \citep[SOHO,][]{Fleck1995}. The LASCO instrument consists of two coronagraphs C2 and C3, with field of views of up to 3~solar radii (hereafter $R_\odot$) and 3 to 32~$R_\odot$ respectively. Furthermore, we present data from the Meudon spectroheliograph in H$\alpha$ (see \url{http://bass2000.obspm.fr/home.php?lang=en}).\\

As a matter of fact, the only two sources of halo CMEs associated with X–class flares in July 2002 were located near the disk center in the active region AR 10030 (\fig{radio}, top panels). They are marked as H27 and H29 in \citet{Bocchialini2018}, but we will call them CME1 and CME2 from here on. The solar source displayed a large sunspot.
\citet{Bocchialini2018} also identified three more CMEs (two partial halos -- P28 and P30 -- and one full halo -- H31) that could have interacted with these two halo CMEs (see \tabl{cme_lasco}, \fig{CME}). 
These flare-CMEs are associated with two sudden storm commencements \citep{Bocchialini2018} with  minimum Dst values of $-17$~nT and $-36$~nT, respectively, as well as solar energetic particles.

We consider a third halo CME (H31-CME3) occurring on July 18 because its high initial speed enables it to interact with CME2 (H29). However, its source region, 
AR 10036,  is located in the South hemisphere close to the limb  (\fig{radio}c).  
The three halo CMEs and the two partial CMEs  are presented in running difference images of LASCO C2 in \fig{radio} (bottom panels). CMEs are defined as halo or partial  CMEs when they \textbf{are visible} in C3 with a halo or not. This depends on their direction and also from their global expansion.
\fig{CME} summarises their characteristics found in the CDAW catalogue (\url{https://cdaw.gsfc.nasa.gov/CME_list/UNIVERSAL/2002_07/univ2002_07.html}). It marks halo/partial halo CMEs,  their launch time, source coordinates, measured height during propagation within the C2 coronagraph, and class of the associated flare.
The P28 and H27 CMEs are very close in space and time. We can assume that they are merging close to the Sun. We keep H27 in our study because H27 is observed at larger distances from the Sun \textbf{compared to} P28 according to the CDAW catalogue. The situation is similar for P30 and H31 and we keep H31. Their interactions close to the Sun can explain why L1 observations are not textbook-case like.

\begin{figure*}[ht!]
\centering
\includegraphics[height=20cm,angle=0]{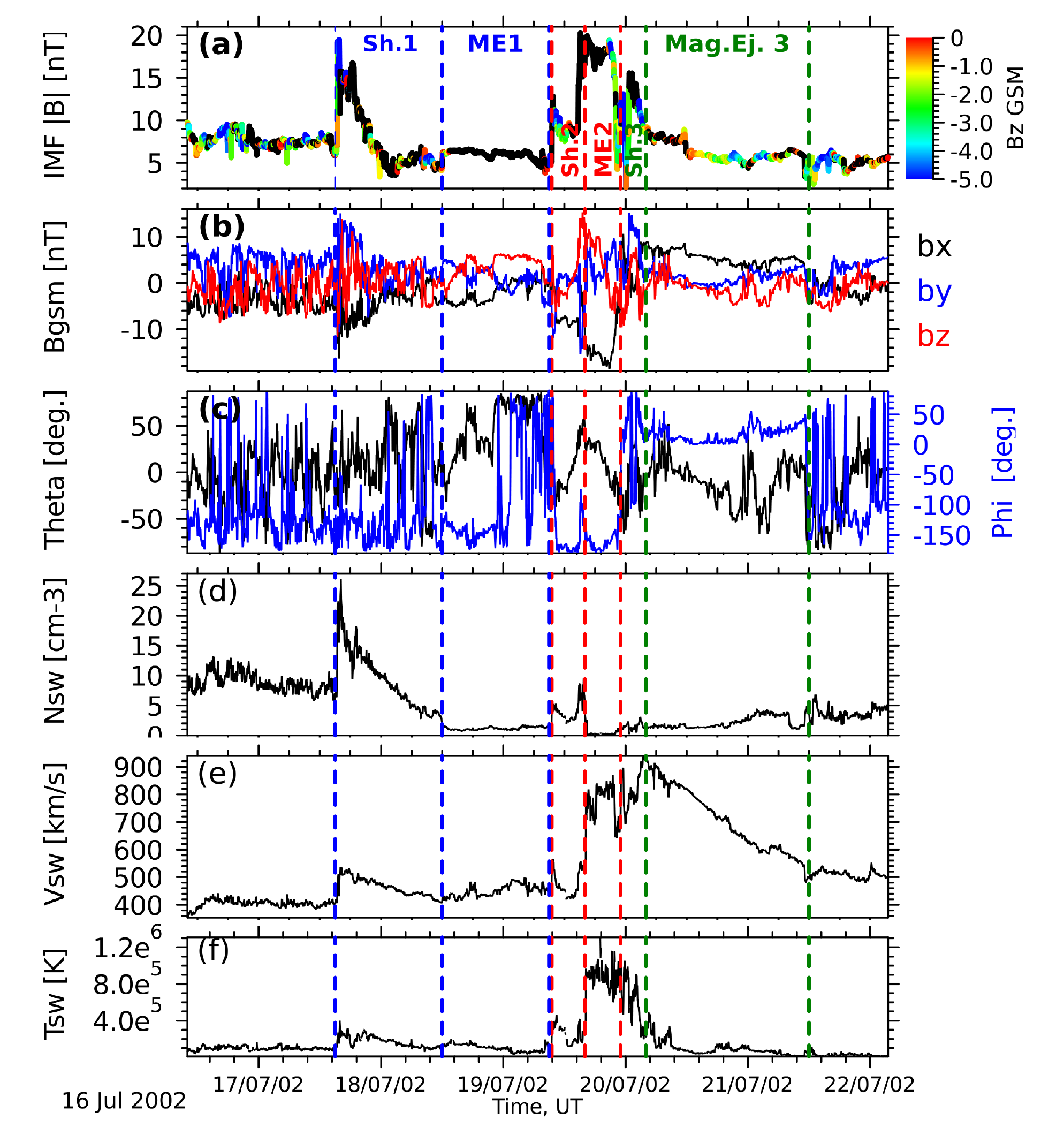}
\caption{
 In situ measurements at L1 with ACE on 16-22 July 2002 for ICME1, ICME2 and ICME3.   (a) Strength, (b) components, and (c) orientation angles of the magnetic field. Proton (d) density, (e) velocity and (f) temperature.
The vertical dashed lines delimit sheaths and magnetic ejecta (MEs, blue for ICME1, red for ICME2, green for ICME3). The color coding of the data points in the top panel is for the North-South B$_{\rm z}$ component (black is for northward orientation).
}
\label{fig_grison1}
\end{figure*}

\subsection{Signatures of the ICMEs at L1}
\label{sect_Obs_Signatures}
Later on, the three CMEs are observed at L1 by the Advanced Composition Explorer \citep[ACE;][]{Stone1998}. 
\fig{grison1} (a,b) presents the IMF magnitude and the three components of the IMF in GSM coordinate system. The coloured parts of the magnitude display the intensity of the southward component of the IMF (black colour means that the IMF is northward at that time). The next panel shows the magnetic field angles to follow its rotation. $\phi_{IMF}$ is the angle in the $(XY)_{GSM}$ plane ($\phi_{IMF}=0$ towards positive X values). $\theta_{IMF}$ is the inclination of the IMF from this plane ($+90^\circ$ for full northward IMF). Finally, \fig{grison1} (d-f) presents the solar wind particle density, velocity and proton temperature ($T_{\rm SW}$), respectively. 

For each ICME we identify a sheath and a ME as defined by \citet{Rouillard2011}.  The sheath is formed by overtaken plasma and magnetic field present earlier on, ahead of the ejecta. This can be either from background solar wind or from a preceding ICME. The interaction transforms the physical properties (e.g. with compression and magnetic reconnection) so much that plasma and magnetic field typically have different properties compared to the region present further away in front (and not yet affected by the interaction).  The identified ME, behind the sheath, designs a magnetic structure more general than a magnetic cloud, while with less strictly defined similar properties (e.g. a less homogeneous rotation of the magnetic field, a less marked temperature decrease, or even the absence of some of the magnetic cloud characteristics). A direct interpretation of these less marked properties is an encounter of the ejecta by the spacecraft at its periphery where all characteristics are expected to be weaker and more perturbed by the encountered surrounding.  
With these definitions, an ICME, with its sheath and ME, is the interplanetary counterpart of a CME as observed with coronagraphs (and/or heliospheric imagers).
ICME1 is associated to the halo CME of July 15, while ICME2 and ICME3 are associated to the two halo CMEs observed on July 18. The overall picture is complex; the physical scenario is a set of three ICMEs in mutual interaction: ICME1 is overtaken by ICME2, which is itself overtaken by ICME3, as detailed hereafter.

The ICME1 observation at L1 starts on July 17 at 16:00 UT with a shock (see \fig{grison1}). The sheath between the shock and the ME displays a strong and fluctuating magnetic field region. We agree with \citet{Richardson2010} with a sheath end time on 18 July 2002 around 12:00 UT and a ME end time on 19 July 2000 around 09:00 UT. The magnetic field is strongly asymmetric.  The speed of the shock front is about 
30 \% to 35 \%
higher than the speed of the background solar wind, which is compatible with the expected ICME speed range \citep[i.e.\ between 18-32\%  according to][]{Temmer2017}.
In the sheath, the density is atypical and decreasing with time, while the velocity profile indicates that the sheath is in expansion, which is typical of slow ICMEs \citep[relative to the front solar wind,][]{Regnault2020}.
The temperature profile has a variation comparable to the velocity profile. Such correlation is generally rather observed and expected in the solar wind \citep[e.g.][and references therein]{Elliott2005,Demoulin2009c}.  

The lower amplitude of the magnetic fluctuations observed after the sheath is typical for a ME, while much larger than within magnetic clouds. This is best shown with the magnetic field angles (see \fig{grison1}c) where a global rotation is present with both angles, despite the presence of large fluctuations.  We notice that the large fluctuations of $\phi$ during the second half of ME1 are \textbf{insignificant} since the magnetic field is dominantly oriented along the z direction.
Other characteristics of ME1 are atypical for isolated ICMEs as follows. Single ICMEs are typically in expansion with a linear decreasing velocity profile and higher magnetic field strength than the background solar wind \citep{Demoulin2009b,Regnault2020,Chane2021}. In case of ICME1, the magnetic field strength in the ME is lower than the much stronger field that is present in front of the sheath. Moreover, ME1 has density values reduced by a factor 5 compared to the solar wind in front of the sheath. The radial velocity component has a non-typical profile with compression that is present starting at about the middle to about 80\% of its size. Indeed, ICME1 is overtaken by ICME2 (see \fig{grison1}).
The proton temperature $T_{\rm SW}$, is only reduced at the rear of the ME1, while $T_{\rm SW}$ is typically lower in MEs by at least a factor of two than its expected value for typical solar wind with same speed \citep[][and references there in]{Elliott2005,Demoulin2009c}. As such, ME1 has atypical properties.
Taking into account the coherence of plasma parameters, in particular density (see \fig{grison1}d), we argue that ME1 is formed only by one CME, while the spacecraft crossed its periphery where the magnetic field is weak and easily modified by interactions with the encountered surroundings along the Sun-Earth travel line.

The \insitu\ profiles of the right side of \fig{grison1} show the two following ICMEs: ICME2 on 19 July 2002 from 09:00 to 23:00 UT, and ICME3 from 20 July 2002 at 04:00 UT to 21 July 2002 at 12:00 UT (see the vertical lines in \fig{grison1}. 
This is justified by the sharp variations of the magnetic field and the plasma parameters.
This interpretation differs from that of \citet{Richardson2010} who consider a single sheath and ICME during that time period. Our interpretation is supported by the different possible solar sources identified for that event in \citet{Bocchialini2018}. 
ICME2 is a small structure (passing through L1 in 14 hours). This is nearly one third of the passage time of ICME1 and ICME3. A plausible explanation is that ICME3 overtakes ICME2, not allowing its radial expansion during its transit from the Sun. Indeed ME2 has nearly a constant velocity (with fluctuations) when observed at L1

The magnetic field strength of ME2 is fluctuating around 18$\;$nT, while a clear rotation of the field components is present (\fig{grison1}c). One noticable characteristic of ME2 is that it has a hot proton temperature that is more typical of sheaths. This is expected to be linked to its non-radial expansion.  The field strength is weaker in ICME3 with a stronger magnetic field in the ICME sheath than in the ME.  The magnetic field orientation is rotating in ME3 while non-monotonously. We notice that important magnetic field fluctuations can be present in the observed central part, even for a magnetic cloud, because the spacecraft trajectory is nearly tangent to the flux rope \citep{Dasso2006}. This could be the origin of the localized changes present in the angles of \fig{grison1}c just after the center of ME3.
  We conclude that ME2 and ME3 have only partly the characteristics of magnetic clouds.  These properties indicate that the spacecraft crossed both ICMEs far from their central region where a cold FR is typically expected.  This is especially true for ME3, as for ME1, while the clear rotation of the magnetic field in ME2 indicates a less extreme crossing.

The plasma density is reduced by one order of magnitude within ME2 with respect to the plasma density found in front of sheath1.
Since the ME2 velocity profile indicates an absence of expansion at L1, this low plasma density is most plausibly intrinsic to its solar origin.
Finally, ICME3 has a typical expansion profile with a low proton temperature and a high velocity in the front with an increase factor of  50\% compared to the solar wind speed  (400~km~s$^{-1}$) ahead of the sheath of ICME1. 

The $B_{\rm z}$ component which is oriented to the north during the time periods of magnetic ejecta 1 and 2 
may explain the weak Dst depressions (not shown here) observed for these events (see \citet{Schmieder2020}).
We note that later on, we compare the observed proton temperature $T_{\rm SW}$ with the plasma temperature $T_{\rm p}$ provided by EUHFORIA simulations, as the latter is assumed to coincide with the proton temperature in the single fluid description that is used during the EUFHORIA MHD simulations.

\section{EUHFORIA simulations}
\label{sect_simulations}


In the following, we aim to first match three-dimensional (3D) MHD EUHFORIA simulations with the ICME observations by ACE at 1$\;$au, and second to find an answer to the following question: is the low density of ME1 and ME2 explained by over-expansion near the Sun or during the Sun--Earth journey, even when no signatures of extreme expansion rate are observed at 1 au from their velocity profiles?

\begin{figure*}[ht!]
\centering
\includegraphics[scale=0.5]{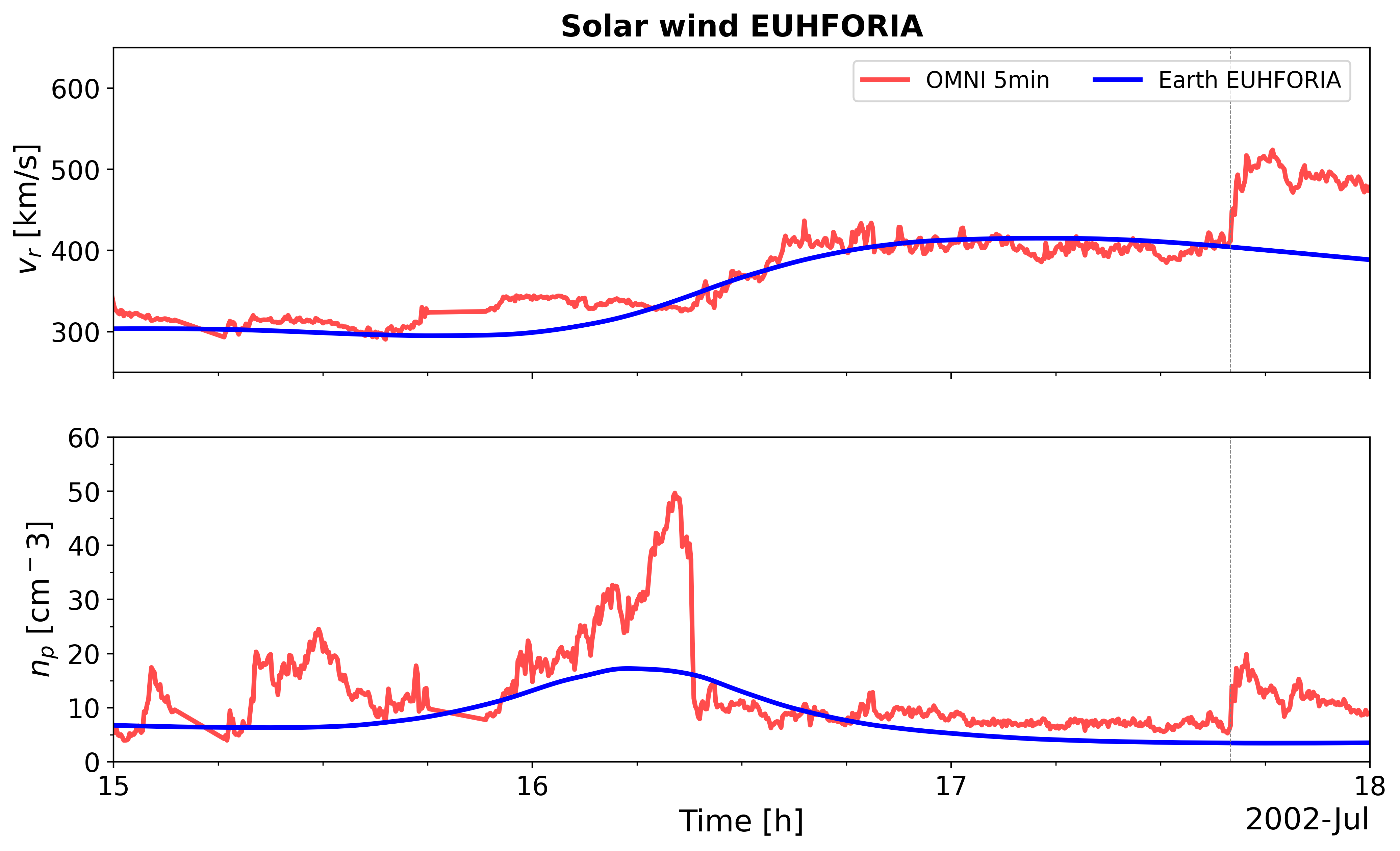}
\caption{EUHFORIA best results of the solar wind computed at L1 (blue curves) between 15 and 18 July 2002 compared to in situ measurements at L1 (red curves).}
\label{fig_euhforia_SW}
\end{figure*}

\subsection{The EUHFORIA model}
\label{sect_simu_EUHFORIA}
Recently, \citet{Pomoell2018} presented EUHFORIA, a space weather forecasting tool for simulating the propagation and evolution of the solar wind and CMEs. EUHFORIA consists of two main parts: a coronal  model and a heliospheric model.

The coronal model is a 3D semi-empirical model that aims to determine the solar wind plasma conditions at 0.1 au viz. the inner radial boundary of the heliospheric model. It uses an approach based on the Wang-Sheeley-Arge model  \citep[WSA,][]{Arge2004}. It requires maps of the photospheric magnetic field, such as those observed by the Global Oscillation Network Group (GONG) of the National Solar Observatory \citep{Harvey1996} or the Mount Wilson Observatory \citep{Howard1976}. More details on the coronal model can be found in \citet{Pomoell2018}. 

The three-dimensional inner-heliospheric model focuses on the dynamics in the inner heliosphere by numerically evolving the ideal MHD equations, including gravity, using a finite volume method together with a constrained transport approach using the Heliocentric Earth EQuatorial (HEEQ) coordinate system. It uses the solar wind plasma conditions that are generated by the coronal model as inner radial boundary conditions. The computational domain of the performed heliospheric simulations extends from 0.1~au to 2~au in the radial direction, $\pm 60\degree$ in latitude and the full $360\degree $ in longitude. 

On top of modelling the background solar wind, the EUHFORIA model allows the modelling of CMEs. CMEs are inserted at the inner radial boundary (0.1~au) as time-dependent conditions. The cone model \citep{Pomoell2018,Scolini2018b} propagates the CME in a simplified way as a hydrodynamic pulse, while the Linear Force-Free (LFF) spheromak model \citep{Verbeke2019b} simulates  a more realistic flux-rope like CME including a magnetic field structure. 
Both models require the same geometric and kinematic CME parameters given by the following: the CME on-set time t$_{\rm CME}$, the radial speed v$_{\rm CME}$, the latitude $\theta_{\rm CME}$ and the longitude $\phi_{\rm CME}$ of the CME propagation direction (in this case its source region), and the CME half-width $\omega_{\rm CME}$.
All of these parameters can potentially be derived from remote-sensing white-light observations of CMEs in the solar corona, especially if observations from multiple viewpoints with multi-filters/spectral data are available. When these are not available, in situ measurements can be used to constrain these parameters after running a variety of cases. 

For both the cone and LFF spheromak models, the CME mass density $\rho_{\rm CME}$ and plasma temperature $T_{\rm CME}$ are taken to be uniform inside the CME with $\rho_{\rm CME} = 1 \times 10^{-18}$~kg~m$^{-3}$ and $T_{\rm CME} = 0.8 \times 10^{6}$~K, similar to \citet{Pomoell2018}. These values are derived from the investigations of the sensitivity of a coronal mass ejection model (ENLIL) to solar input parameters \citep{Falkenberg2010}.
The LFF spheromak model also utilises three additional magnetic parameters, which define the configuration of the inserted magnetic structure: the helicity sign or handedness $H_{\rm CME}$, the tilt angle $\tau_{\rm CME}$, and the toroidal magnetic flux $\phi_{\rm t,CME}$. 
These magnetic parameters can often be constrained using observations, see e.g. \citet{Palmerio2017} or \citet{Scolini2019}.

The work presented here has been obtained by using EUHFORIA version 1.0.4. The computational mesh is uniform in all directions with an angular resolution of $2\degree$ for both latitude and longitude and a total number of 512 cells in the radial direction, leading to a radial resolution of about 0.0037~au, corresponding to 0.798~$R_\odot$. 

\subsection{The solar wind}
\label{sect_simu_solar_wind}

While one can try to simulate the background solar wind and the propagation of the embedded CMEs by using strictly observational input to compare with observational data, here we focus on the opposite: we try to model the observed  data as best as possible. As such, firstly we focus on obtaining a background solar wind with speeds and densities that are similar to the observed quantities. As a result, we have taken the Mount Wilson Observatory magnetogram from Carrington Rotation 1991 as this observational magnetogram provided the best results (see ftp://howard.astro.ucla.edu/pub/obs/\-synoptic\_\-charts/\-fits/\-MP05\_\-5250078\_50\_C1991500\_01.fits).

The standard solar wind model setup from \citet{Pomoell2018} provides a solar wind with too slow speeds and a density peak that is too high compared to observations where the faster solar wind meets the slow solar wind. As such, we have adapted the coronal model inputs to match the observed solar wind by reducing the added speed from $-50.0$~km~s$^{-1}$ to 0.0~km~s$^{-1}$, and by increasing the number density of the fast solar wind from 300~cm$^{-3}$ to 350~cm$^{-3}$ at 0.1~au to provide a better fit to the preceding solar wind observations.  The results at the L1 location are shown in \fig{euhforia_SW} for the time range 15-17 July 2002.

\begin{table*}[ht!]
\centering
\begin{tabular}{lllrrrlllll}
\hline
 CME model & CME \# &  Insertion time  & Speed & Latitude & Longitude & Half width & Tilt & Helicity & Toroidal flux\\                       
           &     & [UT]             & [km~s$^{-1}$] & [$^\circ$] & [$^\circ$] &  [$^\circ$] & [$^\circ$] & sign & [10$^{14}$ Wb]\\
            \hline
            
 Cone       & CME1        & 15 July 23:20 & 1100    &   14                         & 1     & 45 & -- & -- & -- \\
            & CME2        & 18 July 11:10 &  970    &  20      & 15    & 45 & -- & -- & --  \\
            & CME3        & 18 July 20:50 & 2070    & $-10$     & $-40$ & 45 & -- & -- & --  \\
                        \hline 
 Spheromak (Case 1) & CME1 & 15 July 23:20 & 750   & 14    &   1        & 45   & 90    & $+1$ &  0.6\\   
                    & CME2 & 18 July 11:10 & 650   & 20    &  15        & 45   & 90    & $+1$ & 0.6 \\
                    & CME3 & 18 July 20:50 & 1400  & $-10$   & $-40$    & 45   & 90    & $+1$ &0.6\\
                                     \hline
 Spheromak (Case 2) & CME1 & 15 July 23:20  &    700 &      19 & 1      & 30 &180 & $+1$ &0.75\\ 
                   & CME2 & 18 July 11:10  & 935 & 19 & 20              & 45 & 180  & $+1$ & 1\\
                   & CME3 & 18 July 21:00  & 1205 & 19& 20              & 45 & 180 & $-1$ & 1\\
                                     \hline
               \end{tabular}
\caption{Parameters of the three ICMEs at 0.1$\;$au used in the cone and spheromak simulations with EUHFORIA.  
For the cone model, the speeds are taken from LASCO C3 observations at the closest height to $\sim$20~$R_\odot$, as listed in the CDAW CME catalog.
Insertion times are then extrapolated to 21.5~$R_\odot$ assuming a constant speed. For both the cone and LFF spheromak models, the CME mass density $\rho_{\rm CME}$ and plasma temperature $T_{\rm CME}$ are taken to be uniform inside the CME, similar to \citet{Pomoell2018}. For the spheromak model, case~1 parameters are derived using the methodology explained in Section~\ref{sect_simu_ICME} based on LASCO C3 observations at $\sim$20~$R_\odot$. Case~2 uses adapted parameters which have been optimised to get results at 1 au closer to \insitu\ observations, as explained in Section~\ref{sect_simu_ICME}.  
}
\label{tabl_icme}                     %
\end{table*}

\subsection{ICME models}
\label{sect_simu_ICME}
We perform three different simulations: one using the cone model for all three CMEs, and two different setups of the LFF spheromak model with different initial parameters that we will call case 1 and case 2, respectively. As further discussed below, for the cone and spheromak case 1 simulations, the CME initial parameters are derived from LASCO observations in the best way possible after looking at the possible sources of the involved CMEs in the invovled active regions. For the case 2 simulation with the LFF spheromak model, the CME initial parameters are determined by searching for the best match between the observed and model parameters at the L1 location. A summary of the input parameters used for each run is provided in Table~\ref{tabl_icme}.

From LASCO single-viewpoint observations, the observed CME speeds correspond to the speeds $v_{\rm ps}$, i.e. projected in the plane of the sky. They are deduced from LASCO C3 at the height of 20~$R_\odot$ (close to the inner boundary of the heliospheric model), as listed in the CDAW CME catalogue. Given the lack of multi-viewpoint observations for the particular CMEs under study, in this work we perform cone model simulations initialising CMEs with $v_{\rm ps}$, assuming them to be equivalent to the total CME speed in 3D space, i.e. $v_{\rm 3D} \approx v_{\rm ps}$.
The latitude and longitude of insertion correspond to the observed source location (see Table \ref{tabl_cme_lasco}). As the true half-width from halo CMEs is generally difficult to determine, here we have taken a common value of $45^\circ$ for all considered CMEs \citep{Yashiro2004, Robbrecht2009}. 

For the LFF spheromak model, \citet{Scolini2019} showed that a reduced speed, corresponding to the sole radial (translational) speed of the spheromak away from the Sun, should be used to initialise spheromak CMEs in EUHFORIA, to avoid super-expansion effects (and earlier arrival time predictions at Earth) due to the initial pressure imbalance between highly-magnetised spheromak structures and the surrounding solar wind medium. 
To address this issue, and due to the lack of observations of the radial speed for halo CMEs, in this work we derive the CME radial speed for EUHFORIA spheromak simulation case 1 using a combination of two approaches: (1) a method applying the relations by \citet{Schwenn2005} to spheromak CMEs observed from single viewpoints as developed by \citet{Scolini2019}, here used in combination to the the assumption that $v_{\rm 3D} \approx v_{\rm ps}$ as done for cone model simulations and explained above; and (2) a direct application of the method based on previous works by \citet{Schwenn2005}, under the assumption that $v_{\rm ps} \approx v_{\rm exp}$.
The authors showed that the radial speed, $v_{\rm rad}$, of the leading edge is statistically related to the lateral expansion speed between the lateral sides, $v_{\rm exp}$, for CMEs launched near the solar limb. Using 57 limb CMEs, they found the linear relationship $v_{\rm rad} = 0.88\, v_{\rm exp}$ with a correlation coefficient of 0.86 (see Figure 7 therein). For halo CMEs, while $v_{\rm rad}$ is not observable due to the lack of information in the direction perpendicular to the plane of the sky, $v_{\rm exp}$ is still measurable as $v_{\rm exp} \approx v_{\rm ps}$ and was shown to be nearly isotropic. Therefore, assuming that the 57 limb CMEs used are typical of the whole set of CMEs, the measure of $v_{\rm exp}$ allows to estimate $v_{\rm rad}$ for specific halo CMEs. As a result, for the LFF spheromak model in case 1, we calculate the radial insertion speed as the mean of two values: 
(1) $v_{\rm rad} = 0.47\, v_{\rm 3D} = 0.47\, v_{\rm ps}$, where the LASCO speed value $v_{\rm ps}$ is considered as equivalent to the 3D value $v_{\rm 3D}$.
And (2) $v_{\rm rad} = 0.88 v_{\rm exp} = 0.88 v_{\rm ps}$, where the LASCO speed value $v_{\rm ps}$ is considered as equivalent to the expansion speed $v_{\rm exp}$. 
The resulting relation used to derive the initial speed for EUHFORIA simulation case 1 from LASCO C3 observations is $v_{\rm rad} = 0.68\, v_{\rm ps}$. Such a hybrid method was chosen after performing additional simulations (not shown) using methods (1) and (2) separately, and assessing that the arrival time of the ICMEs at 1~au were off by $-4$ and +10 hours, respectively. For this reason, we have settled on a hybrid approach which overall proved to perform best in reproducing the ICME kinematics for the particular cases investigated in this study, resulting in a late arrival of only 2 hours.

For the parameters concerning the helicity sign, the tilt angle and the magnetic flux, we have extremely limited observational data to constrain their values. The helicity sign in simulation case 1 is fixed to $\HCME = +1$ (i.e. right-handed) for all CMEs, consistent with the helicity sign inferred from in situ observations of ME2 exhibiting a North-East-South rotation (Figure~\ref{fig_grison1}). 
The tilt angle is chosen equal to $\tauCME =90^\circ$. 
The toroidal magnetic flux is fixed to $\phitCME = 0.6 \times 10^{14}$~Wb. 

For case 2, we explore the space of parameters for each relevant physical quantity (i.e. longitude, latitude, speed, half width, helicity, tilt, and toroidal magnetic flux),
and finally present the best results where the arrival time, the density and velocity profiles of the simulations are in closer agreement with the observations (see Figure \ref{fig_euhforia_SW}). Here, the helicity sign is fixed to $\HCME = +1$ (i.e. right-handed) for all CMEs except for CME3, where the helicity sign is taken as $\HCME =-1$ (left-handed). We note that the simulation results for a similar run as case 2 with all helicities fixed to $\HCME=+1$ provides very similar results for the analysis performed later, with only minor differences in the predicted arrival times and differences in polarity of the magnetic field components. We refer the reader to \citet{Taubenschuss2010} for work on the effect of magnetic handedness on cloud propagation using 2.5D MHD simulations.
The tilt angle is chosen equal to $\tauCME =180^\circ$ and finally, 
the toroidal magnetic flux is set to $\phitCME = 0.75 \times 10^{14}$, $10^{14}$ and $10^{14}$~Wb for the 3 CMEs, respectively. 

\subsection{Simulation results at 1~au}
\label{sect_simu_Simulation}

\begin{figure*}[ht!]
  \centering
\includegraphics[scale=0.3]{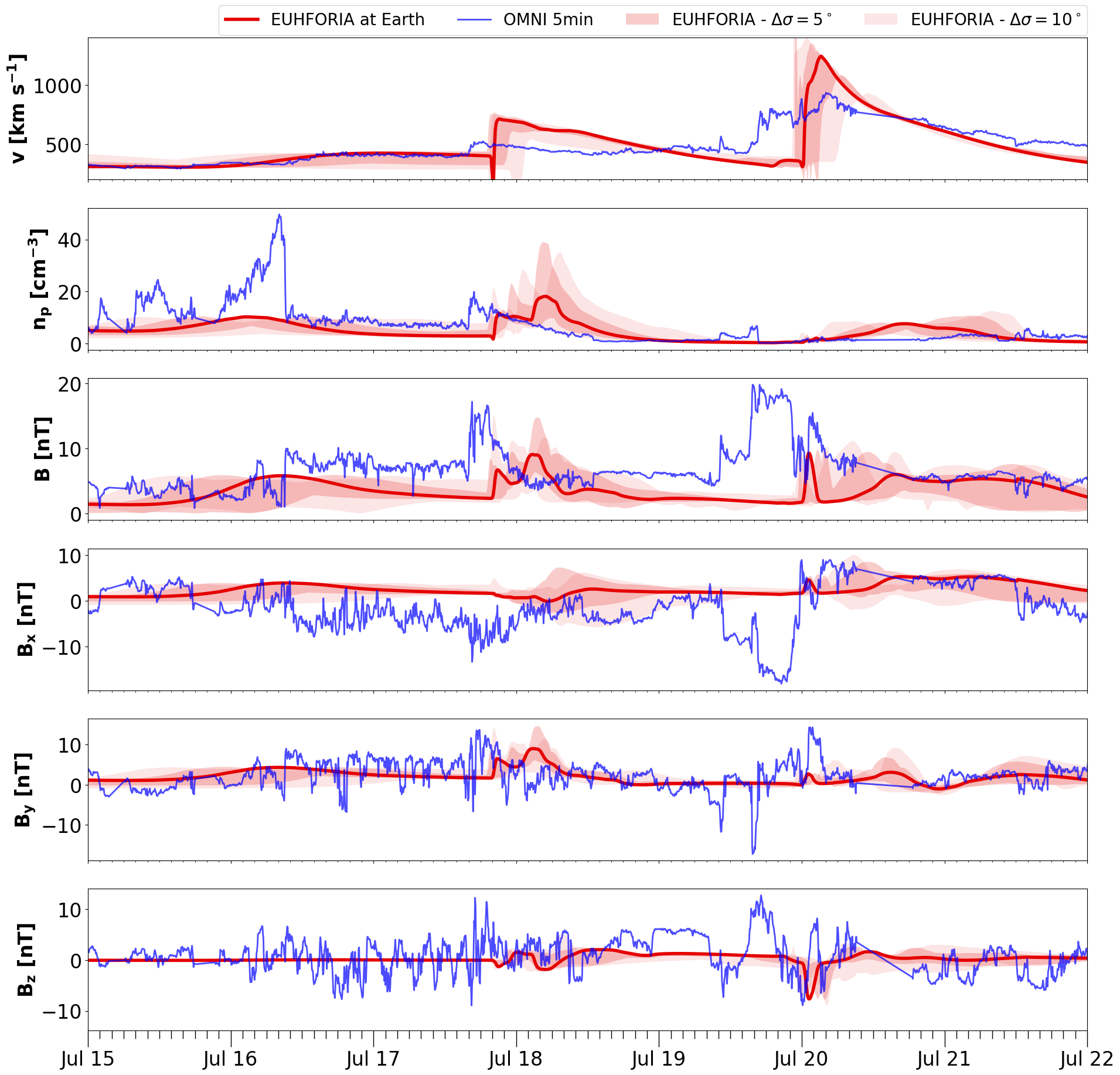}
\caption{EUHFORIA results for the cone model run with three ICMEs (red curve) compared to the OMNI data at L1 (blue curve). The cone model parameters are provided in \tabl{icme}. The dark red and light red shaded areas show the maximum variation of EUHFORIA
predictions at positions separated by $\Delta\sigma = 5\degree$ and $\Delta\sigma = 10\degree$ in longitude and/or latitude from Earth. From top to bottom: speed ($v$), proton number density ($n_{\rm p}$), magnetic field magnitude ($B$) and
magnetic field components in GSE coordinates ($B_{\rm x}$, $B_{\rm y}$, $B_{\rm z}$). We note that OMNI data has a gap around 20 July 2002 at 12:00 UT. We refer the reader to Figure \ref{fig_grison1} for ACE observations.}
\label{fig_EU_fit3}
\end{figure*}

We first present results from the cone model in \fig{EU_fit3}. The arrival times of the ICMEs are late by one hour for the first ICME and four hours for the second one, while the speeds are too high. The third CME is too deviated from the Sun--Earth direction so never reaches Earth's location, nor any of the virtual spacecraft. It is worthwhile noting that while the CMEs are arriving late, their arrival speeds are also much higher than the observed speeds. Using the inputs from the cone model simulation run, as given in Table~\ref{tabl_icme} for the drag-based DBM model \citep{Dumbovic2018}, we obtain a similar trend for both arrival times and arrival speeds. As such, it will be hard to match observations with the LFF spheromak model as well. 
\begin{figure*}[ht!]
\centering
 {{\includegraphics[scale=0.14]{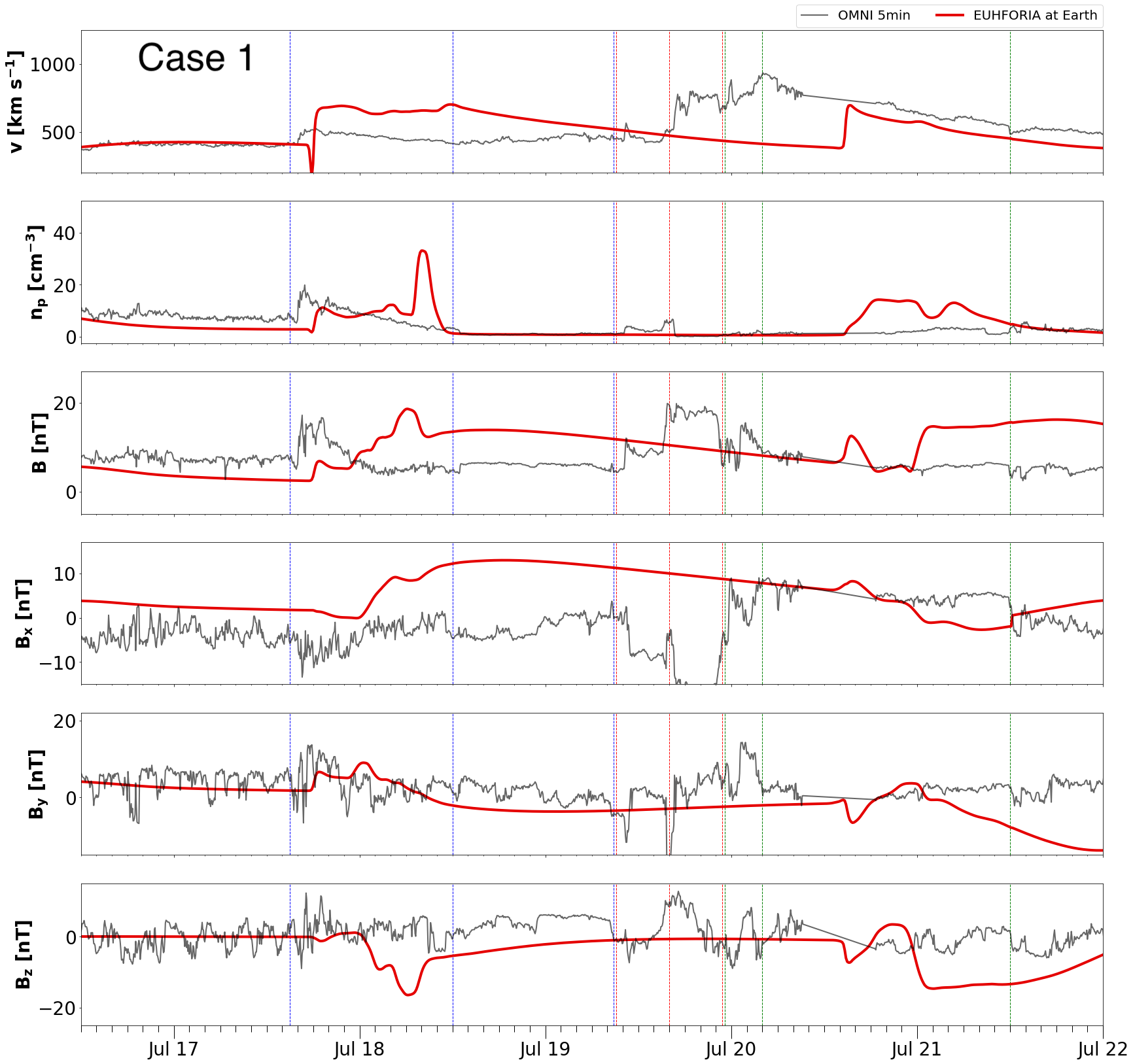} }}
    \qquad
        {{\includegraphics[scale=0.14]{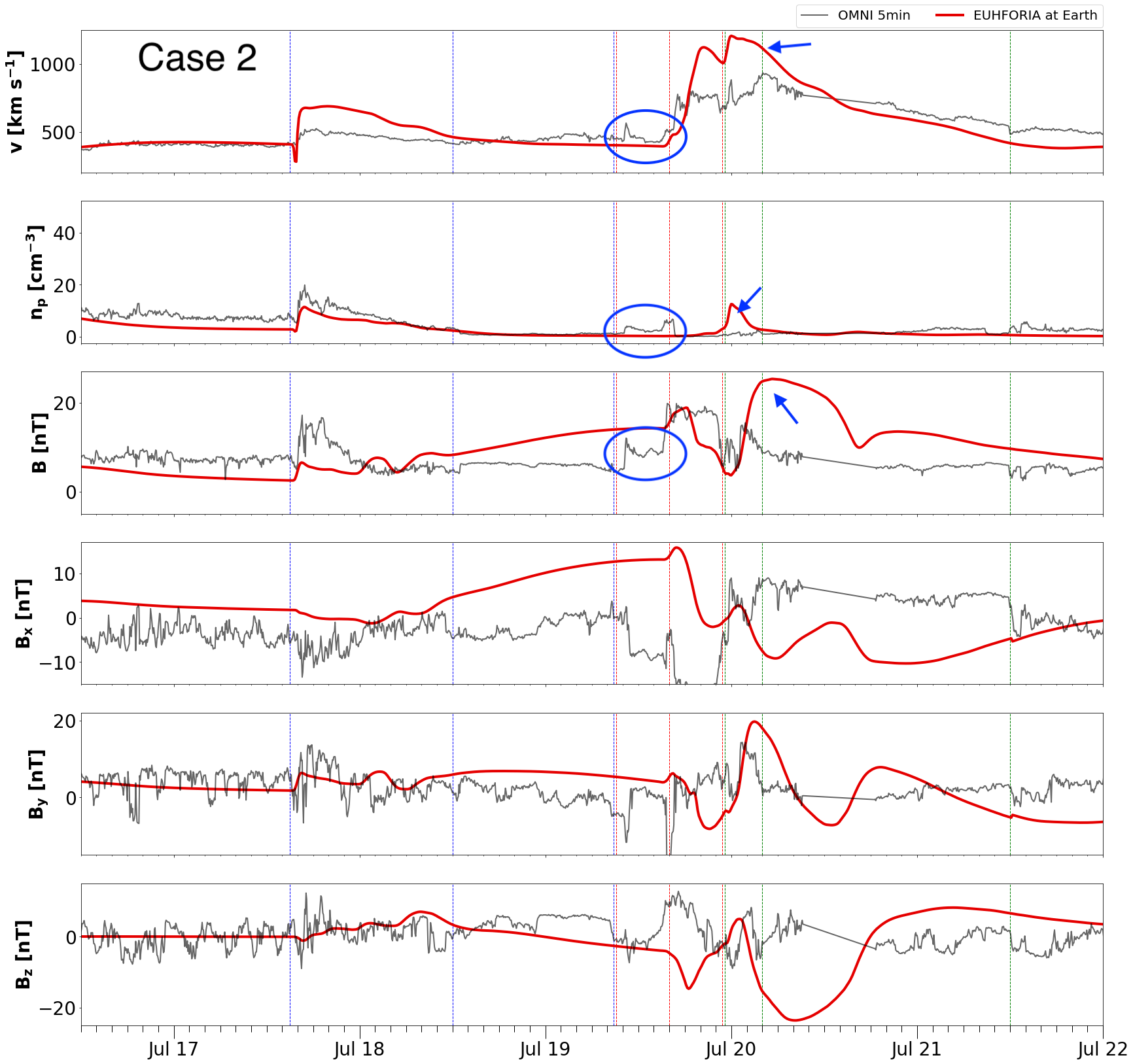} }}%
\caption{EUHFORIA results at L1 location of the modelling the three ICMEs using the LFF spheromak model: radial velocity, proton density, magnetic field strength and components $B_{\rm x}$, $B_{\rm y}$, $B_{\rm z}$  (red curves) compared with the {\it in situ} observations (grey curve). Case 1 and 2 can be found in the left and right panels, respectively. The three observed ICMEs are indicated by the vertical dashed lines that delimit sheaths and magnetic ejecta (blue for ICME1, red for ICME2, green for ICME3 similar as in Figure \ref{fig_grison1}).
Some peaks in the speed, density and B profiles are observed within the sheath of ICME2 (blue ellipses in the right panels), indicating the existence of a compression which does not appear in the simulation. 
On the right panels, the blue arrows indicate the main difference between the observations and the simulations. 
Indeed, strong peaks in velocity, density and field strength are formed in the simulation due to ICME3 overtaking ICME2, but their counterparts within \insitu\ data are not present.}
\label{fig_Case1_case2_1}
\end{figure*}

The results of the simulations performed in the two LFF spheromak model cases 
are presented in \fig{Case1_case2_1}, and we compare the results for the radial velocity, proton density, and magnetic field strength and components $B_{\rm x}$, $B_{\rm y}$, $B_{\rm z}$ with observations.
For case 1, \fig{Case1_case2_1} left side, the arrival time of ICME1\simuone\ exceeds by one hour the observations, while ICME2\simuone\ is behind by 18 hours (19 July 2002 at 17:00 UT to 20 July 2002 at 13:00 UT). ICME3\simuone\  does not encounter the spacecraft in case 1.
We notice significant differences between the results of case 1 and observations. These motivate us to perform a partial exploration of the space parameters in order to have an arrival time closer to the one observed and ICME3\simu\ overtaking ICME2\simu\ at L1. The best achievement, among our series of tests is shown as case 2.

The most convincing achievement of the simulations is the formation of the low density region inside ME1\simu\ in both of the cases.  
Case 2 represents better the observed density in ICME1\obs\ with a decreasing density in the sheath almost as observed (\fig{Case1_case2_1}, right side).
Finally, the observed low density inside  ME2\simutwo\ is not recovered in both cases. This suggests that the spheromak model is not efficient in reconstructing the observed structure. The low density could also be due to characteristics of the CME close to the Sun, occurring below the inner boundary of the heliospheric model.

The global comparison of observed and simulated ICMEs shows that case 2 is closer to observations.
Nevertheless for case 2,  there are significant differences between observations and the simulations as follows. The velocity is still faster in the first sheath, in ME2\simutwo\  and in the sheath of ICME3\simutwo\  than in observations. 
In the sheath of ICME2\obs\ tiny peaks are observed in the speed, density and B profiles (blue ellipses in the right panels), indicating  the existence of an observed compression which does not appear in the simulation (partly due to a much coarser spatial resolution). Rather a too strong peak of the speed is present in ME2\simutwo.
Within the sheath of ICME3\simutwo , velocity and density are also well enhanced compared to observations. All these enhancements indicate a too strong interaction between ICME2\simutwo and ICME3\simutwo in comparison to observations. This can be an effect of a bias due to the inserted speed. Next, the velocity and the density in ME3\simutwo\ is about comparable to the observed ones. However, the magnetic field strength is too large in front of ICME3\simutwo.  This implies a too strong magnetic pressure which compress the structures in front. 

Using different parameters (tilt and helicity sign values), we do not succeed to get the computed magnetic field components closer to the observations than in case 2 ( see Figure \ref{fig_Case1_case2_1}, right side) .
The difference in the components $B_{\rm x}$ of the 3 ICMEs\simutwo\ is due mainly to their differences in longitude. The last ICME\simutwo\  has an inverse sign of helicity than the two others which leads to an inverse $B_{\rm y}$ and $B_{\rm z}$ profiles.  Finally, the magnetic field components are very different from the observations in both cases and difficult to explain. Then, in \sect{evolution} we analyse the evolution of the magnetic field and the plasma parameters during the travel from the Sun.

\subsection{Flux rope expansion rates}
\label{sect_evol_Expansion}

The ICME\simu\  speed profiles show evidences of expansion (see \fig{Case1_case2_1}) while the CMEs\simu\  were launched at $0.1$ au with a uniform velocity. Then, below we quantify the expansion at 1 au both for observations and simulations.

The  normalised  rate of  expansion is estimated from \insitu\ data with the formula \citep{Gulisano2010} 
  \BE  \label{eq_zeta}
  \zeta = \frac{v_f-v_e }{\Delta t} \, \frac{D}{ v_c^2}\,, 
  \EE
where $v_f$ and $v_e$ are the velocity at the front and at the end of the ME, respectively, $v_c = (v_f+v_e)/2$, $\Delta  t $ is the difference of times between the front and the end of the ME and $D$ the distance between the centre of the ME and the Sun.
$(v_f-v_e)/\Delta t$ represents the mean slope of the velocity profile as determined e.g. by a linear fit of $v(t)$. 

The observed velocity inside ME1\obs\ is fluctuating, with a global trend of an increasing velocity (left panel of \fig{grison1}), so ME1\obs\ is in compression with $\zeta_{ME1,obs} \approx -0.55$.
This is in contrast with EUHFORIA, where ${\bf \zeta_{simu}}$ is $\approx 1.05$ in ME1\simu\ for both cases.  This is a slightly faster expansion than the typical $\zeta$ values $0.8 \pm 0.2$ found in MCs at 1 au \citep{Demoulin2008} and to $0.9 \pm 0.2$ found in the inner heliosphere with HELIOS spacecraft \citep{Gulisano2010}.
  
Next, ME2 is compressed both in observations and in the case 2 by the overtaking ICME3 (see \fig{Case1_case2_1}). Its velocity is fluctuating around a constant value for observations ($ \zeta_{ME2,obs} \approx 0$) while it is very strongly compressed in the simulation, with $ \zeta_{ME2,simu} \approx -6$, and even the velocity profile is strongly linear so a differential compression is present.

Finally, the observed velocity profile is almost linear within ME3\obs\ (right panel of \fig{grison1}) indicating a self similar expansion in the radial direction. 
The value $\zeta_{ME3,obs} \approx 0.8$ for ME3\obs\ is close to the expected value. A much larger expansion, $\zeta_{ME3,simu,2} \approx 1.8$, is found with EUHFORIA for ME3\simutwo .  
To understand such difference it is required to analyse how the flux rope evolves from the injection distance ($0.1$ au) to 1 au (see next section).

The observed profiles of ICME2 and ICME3 are similar to the ones for two ICMEs in-situ observed during May 2005 \citep[see Figure 1 of][]{Dasso2009}, where from a deep analysis using combined observations, was concluded that the strong distortion of the velocity profile of the first ICME was a consequence of important ICME-ICME interaction. Here the physical scenario is even more complex (a possible interaction between three ICMEs) and will be analyzed later on with EUHFORIA numerical simulations.


\section{What happens between 0.1~au and 1~au?}
\label{sect_evolution}

\subsection{Background solar wind}
\label{sect_evol_Background}
\begin{figure*}[ht!]
\centering
 \includegraphics[scale=0.55]{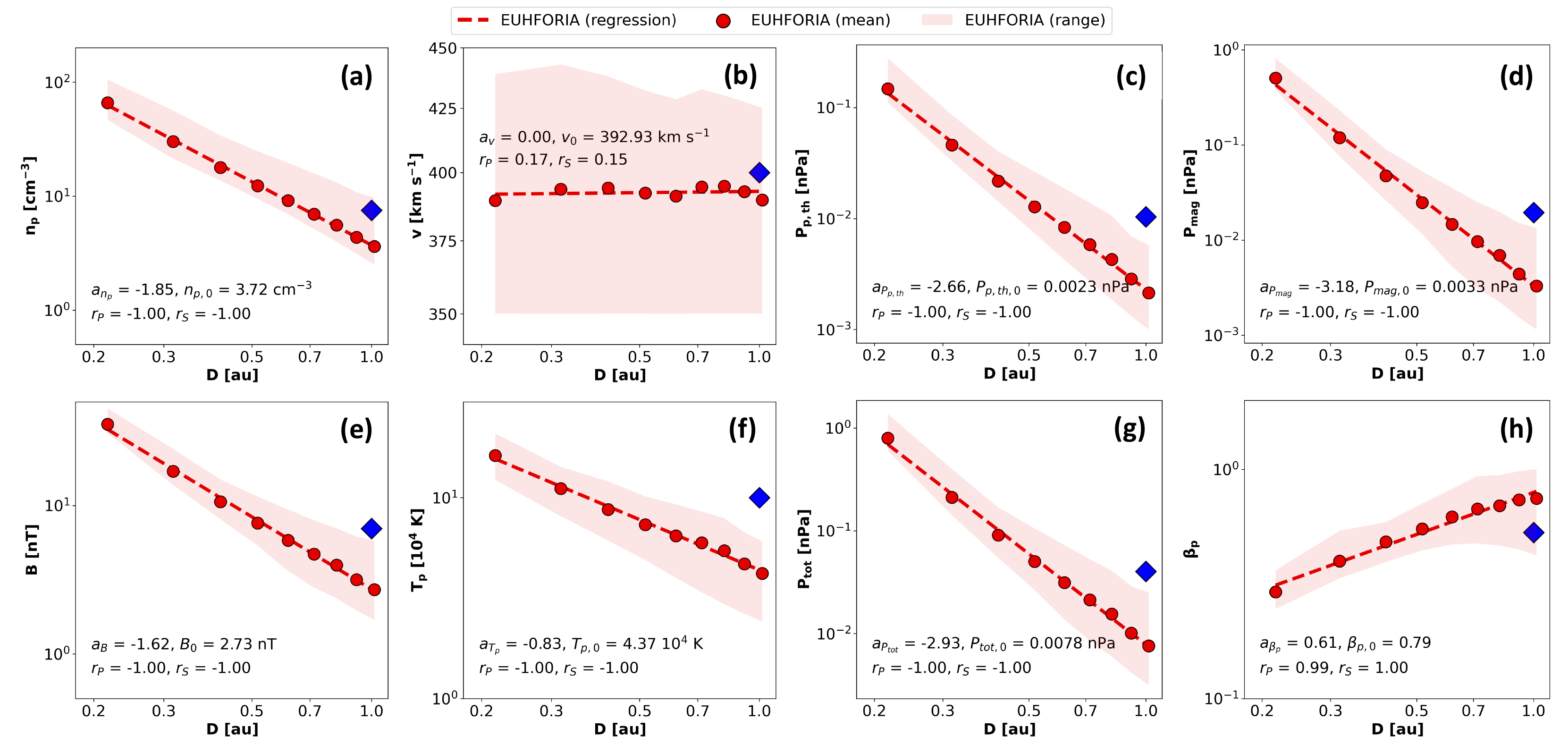}  
 \caption{Ambient solar wind from EUHFORIA: radial dependence of the mean solar wind parameters extracted at virtual spacecraft located along the Sun–Earth line (red dots). 
 These values have been obtained by considering solar wind streams between speeds of 350 and 450 km~s$^{-1}$, i.e. consistent with the solar wind speed preceding ICME1 as visible in Figure \ref{fig_euhforia_SW}.
 (a) proton number density $n_{\rm p}$, 
 (b) plasma speed $v$, 
 (c): proton thermal pressure $P_{\rm p}$,
 (d): magnetic pressure $P_{\rm mag}$, 
 (e): magnetic field $B$, 
 (f): plasma temperature $T_{\rm p}$, 
 (g): total pressure $P_{\rm tot}$,
 (h): proton $\beta$.  
The red shaded areas show the maximum variation of EUHFORIA predictions as a function of the heliocentric distance. The results from the fitting of the mean values are indicated as dashed red lines.
Observed average values from ACE in the 24~hours prior to the arrival of ICME1 are shown as blue diamonds. 
}
\label{fig_evolution}
\end{figure*}

One of the main causes of the flux rope expansion is the decrease of the total pressure ($P_{\rm tot}$) of the ambient solar wind through which they travel \citep{Demoulin2009}. These authors modelled the evolution of a flux rope as a series of force free field states with a pressure balance between the ambient solar wind and the flux rope boundary.  The main driver of the  expansion is the variation of the SW total pressure with solar distance. 
It is well approximated by a power law of the solar distance $D$: $P_{\rm tot}(D)=P_{\rm tot}(D_0)\,  (D/D_0)^{a_{\rm Ptot}}$ where $D_0$ is a reference distance (e.g. 1 au).  This pressure decrease induced a power law increase of the flux rope size $S$ as $S = S_0\,  (D/D_0)^\zeta$ where $\zeta$ is the expansion rate. 
Assuming conservation of the magnetic flux within the FR, \citet{Demoulin2009} found $\zeta = -a_{\rm Ptot/4} \pm 0.2 $ for different twist profiles.

Next, we characterise the properties of the background solar wind in function of the heliodistance $D$. \fig{evolution} shows the evolution of different properties of the background solar wind simulated by EUHFORIA from 0.2~au to 1~au, in a similar format to what is shown in Figures~5 and 6 of \citet{Scolini2021}. 
Specifically, we track all parcels of background solar wind characterised by a speed equal to $400 \pm 50$~km~s$^{-1}$, i.e. consistent with the solar wind speed preceding ICME1 at 1~au as visible in \fig{grison1} and Figure \ref{fig_euhforia_SW}. 
Specifically, the tracking was then done as follows: at each virtual spacecraft/heliocentric distance considered, we scanned the speed time series over the full simulated period (July 15 to July 22), and selected all sub-periods where the speed was within the range $400 \pm 50$~km~s$^{-1}$. Other solar wind properties (i.e. plasma density, temperature, magnetic field, and pressures) were then extracted across these same sub-periods. The results of this analysis in terms of the average/spread in each parameter, for each heliocentric distance, are reported in \fig{evolution}, which shows (from the upper left to the lower bottom panels): the proton density, the bulk velocity, the thermal pressure, the magnetic pressure, the interplanetary magnetic field strength, the plasma temperature, the total pressure, and the plasma beta.

The theoretical expectations for these physical quantities were recently revised in Section 3.1 of \citet{Scolini2021}. 
As expected, EUHFORIA simulations provided an almost constant radial solar wind velocity up to $\sim$ 1~au (\fig{evolution}b).  
We note that this result, which is a direct consequence of the selection of solar wind parcels with speed equal to $400 \pm 50$~km~s$^{-1}$ at all heliodistances, may not necessarily correspond to following the same blobs of plasma during propagation to 1~au. 
This detail and the consequences are further discussed below in relation to the modelled density radial scaling, mass flux conservation, and implications for the expected ME expansion rate.

The profile of $n_{\rm p}(D)$ shows a global decrease such as $n_{\rm p}(D) \propto  D^{-1.85}$, so with a lower decrease with respect to the expected 2D expansion for a constant radial bulk velocity ($n_{\rm p}(D) \propto D^{-2}$).
Strictly speaking, this result does not satisfy the mass flux conservation ($n_{\rm p}\, v\, D^2$ preserved), which would require, for $v(D) \propto D^{-0.00}$ found above, a density radial decay as $n_{\rm p}(D) \propto D^{-2}$. By contrast, in \citet{Scolini2021} the density slope was $\propto D^{-2.07}$, i.e. slightly steeper than the expected behaviour. The apparent violation of the mass flux conservation in the current and previous works may be due to multiple reasons, including non-radial expansion and the fact that, as pointed out before, tracking all solar wind parcels with speed equal to $400 \pm 50$~km~s$^{-1}$ at different heliodistances may not necessarily correspond to following the same blobs of plasma during propagation to 1~au. As individual solar wind parcels propagate towards larger $D$, they are continuously interacting with the surrounding wind, which can alter their properties such as speed and density. For example, interactions among solar wind parcels with different speeds can in turn lead to, e.g. acceleration of slower solar wind which got pushed and compressed by a faster wind coming from behind. Or oppositely, to fast solar wind that got compressed and decelerated after taking over a slower stream ahead.

However, we note that setting $n_{\rm p}(D) \propto D^{-2}$ and assuming no change on the scaling of the plasma temperature would change the scaling of the plasma pressure from $P_{\rm tot}(D) \propto D^{-2.9}$ to $P_{\rm tot}(D) \propto D^{-{\bf  3.05}}$. 
Yet, this is expected to have only a minor effect on the expected expansion rate of MEs ($\zeta$, Equation \ref{eq_zeta}), which is primarily defined via the total pressure as $\zeta \approx - a_{\rm Ptot}/4$. 
The expected change in this case would be from $\zeta \approx 0.73$ to $\zeta \approx  0.77$. 
Based on these arguments, we conclude that the deviation of the density scaling from the expected behaviour most likely has no significant effect on the resulting ME expansion rate obtained from simulations. Yet, further investigations on the apparent violation of mass flux conservation in MHD simulations may be important to validate the solar wind and CME modelling in EUHFORIA, and are left for future studies. 

The plasma temperature decreases as $T_{\rm p} \propto D^{-0.83}$.  This temperature decrease is less than modelled in \citet{Scolini2021}, where the slope is $-0.96$.  However, the obtained plasma temperature decrease is close to the observed values \citep{Perrone2019a}, and also less important than for an adiabatic solar wind expansion ($T_{\rm p} \propto D^{-1.3}$), as expected due to the solar wind heating. Next, the thermal proton pressure simulated by EUHFORIA follows approximately the law $P_{\rm p,th}(D) \propto D^{-2.66}$. This is coherent with the product of the above power laws, i.e. $n_{\rm p}\, T_{\rm p} \propto D^{-2.68}$.

The magnetic field strength scales as $B(D) \propto D^{-1.62}$, which provides a decay as estimated recently from \insitu\ measurements ($B(D) \propto D^{-1.6}$, \citep{Hellinger2011,Perrone2019a}). The magnetic pressure decay with heliodistance is $P_{\rm mag}(D) \propto D^{-3.18}$, close to the square of $B(D)$ power law.

Next, the total pressure (thermal plus magnetic) simulated in EUHFORIA results as $P_{\rm tot}(D) \propto D^{-2.93}$, which is in between the laws for $P_{\rm p,th}(D)$ and $P_{\rm mag}(D)$, as expected. Finally, the plasma $\beta$ slightly increases with $D$ since $P_{\rm p,th}$ decreases slightly less rapidly than $P_{\rm mag}$ with $D$.

In present numerical simulations, $a_{\rm Ptot} = -2.93$ (see \fig{evolution}). 
Thus, if the three ICMEs studied in this paper would travel in a clean ambient solar wind, and expand in force balance with the total pressure of the surrounding solar wind, the expected expansion rate is $\zeta \approx - a_{\rm Ptot}/4 \approx 0.73$.  This is different than the variety of $\zeta $ values measured at 1 au in the three simulated ICMEs (see \sect{evol_Expansion}),  so that a physical analysis of the numerical results is needed.

\subsection{Overview of 3D spheromak propagation}
\label{sect_evol_Overview}

\begin{figure*}[ht!]
\centering
\includegraphics[scale=0.73]{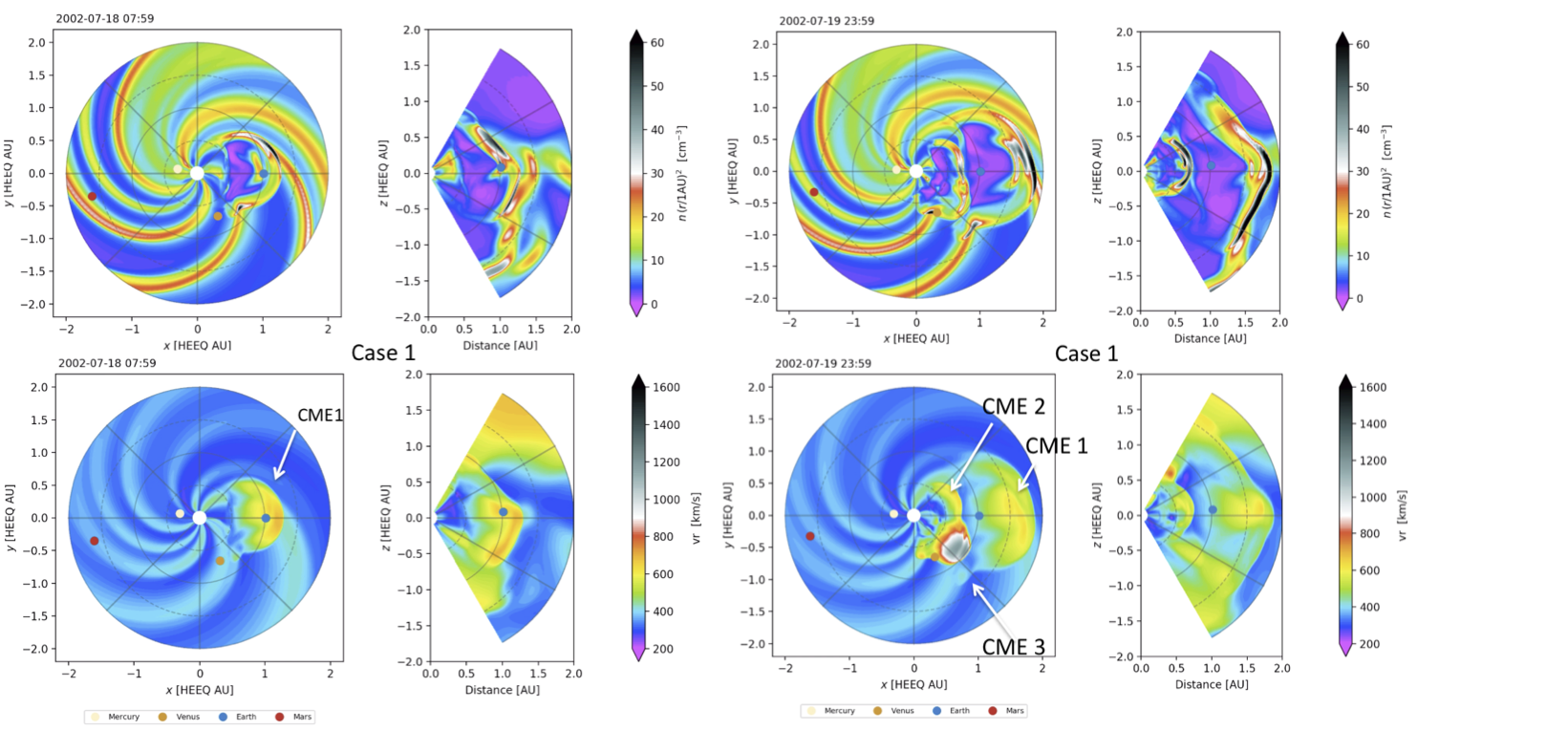}
\caption{
Overview of EUHFORIA results of the three ICMEs modeled with the LFF spheromak model for case 1:  scaled plasma number density (top panels) and radial velocity (bottom panels). The four left panels show ICME1 on 18 July 2002 at 07:59 UT, the four right panels show the three ICMEs on 19 July 2002 at 23:59 UT.  At both times, both an equatorial x-y plane (left) and a meridional distance-z (or x-z) plane (right) are shown.   The full movies are included as supplementary material.
}
\label{fig_EU_fit5}
\end{figure*}

\begin{figure*}[ht!]
\centering
\includegraphics[scale=0.73]{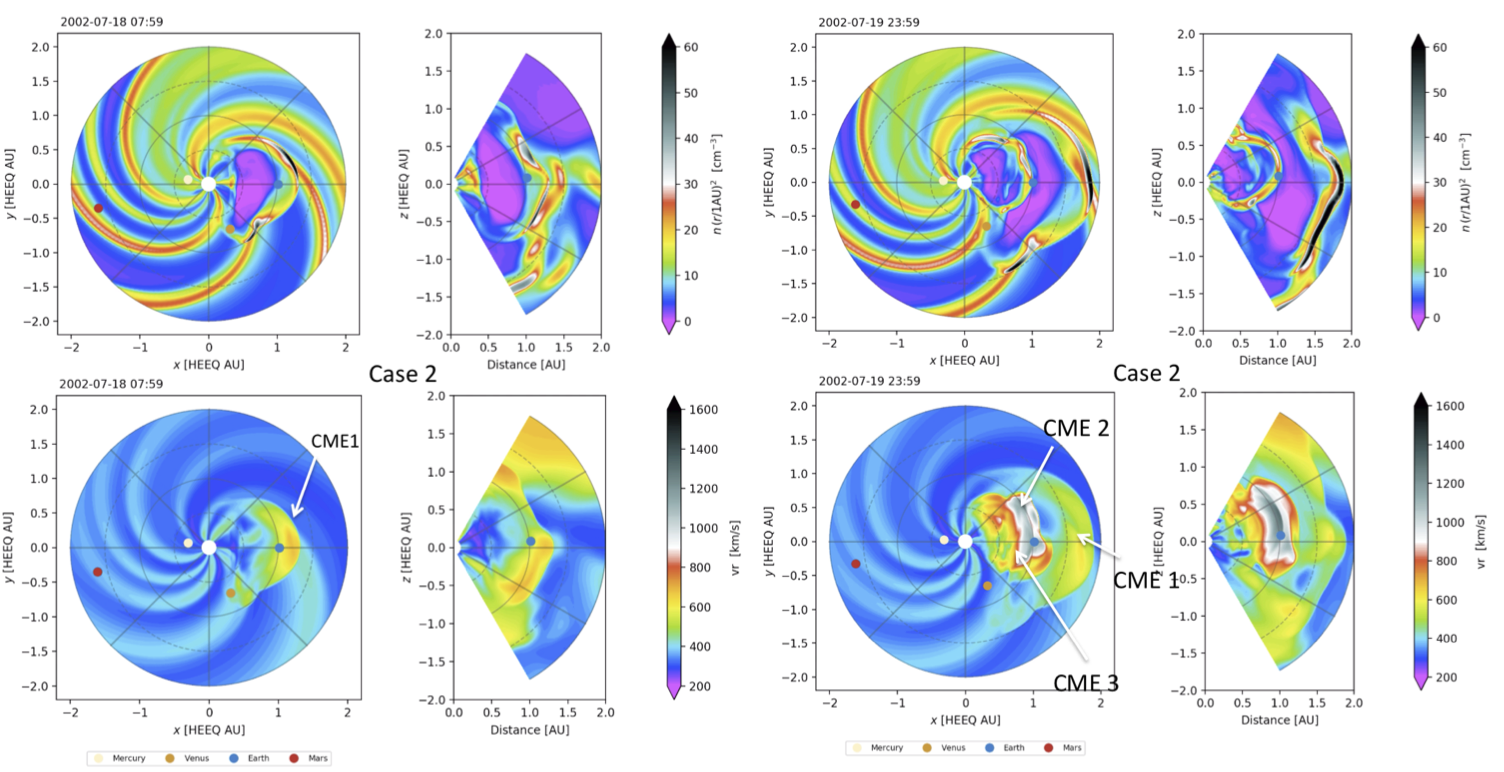}
\caption{
Overview of EUHFORIA results of the three ICMEs modeled with the LFF spheromak model for case 2. The drawing conventions and the selected times are the same as in \fig{EU_fit5}.
The full movies are included as supplementary material.
}
\label{fig_EU_fit6}
\end{figure*}

\figs{EU_fit5}{EU_fit6} show movie snapshots of the scaled number density (scaled by distance squared; top panels) and velocity (bottom panels) showing the propagation and successive interaction of the three ICMEs within two planar cuts (x-y, x-z) for case 1 and case 2, respectively. In case 1, ICME3 is not front sided and only its flank is arriving at the Earth (blue dot). This implies an arrival time too late by 18 hours. In case 2, the three ICMEs are front side. 

For both cases, in the equatorial plane, ICME1\simu\ strongest velocities stays arch shaped with a nearly circular front shape which extends progressively with a nearly constant angular width as seen from the Sun.
In front, an arch shaped strong density enhancement is present. It is formed of compressed solar wind (the sheath).  The angular lateral extension of this density arc is slightly increasing as ICME1\simu\ propagates away from the Sun.

A slower velocity is present as one goes further away behind the fast front of ICME1\simu. This characterises a strong radial expansion of the ME\simu. Together with the orthoradial expansion, which is about proportional to the distance $D$, this implies a 3D expansion of the ME with a strong decrease of density as outlined by the magenta region in the top panels.

Later on, \figs{EU_fit5}{EU_fit6} and the associated movies illustrate the interactions of the ICMEs as ICME2\simu\  and ICME3\simu\ are launched in comparable directions (taking into account their angular widths).   The interaction is more important in case 2 with ICME3\simutwo\ strongly overtaking and deforming ICME2\simutwo .  The same occurs more mildly with ICME2\simutwo\ overtaking and deforming the rear part of ICME1\simutwo .

\subsection{Simulated \insitu\ evolution between 0.1 and 1~au}
\label{sect_evol_Evolution}

\begin{figure*}[ht!]
\centering
 \includegraphics[scale=0.40]{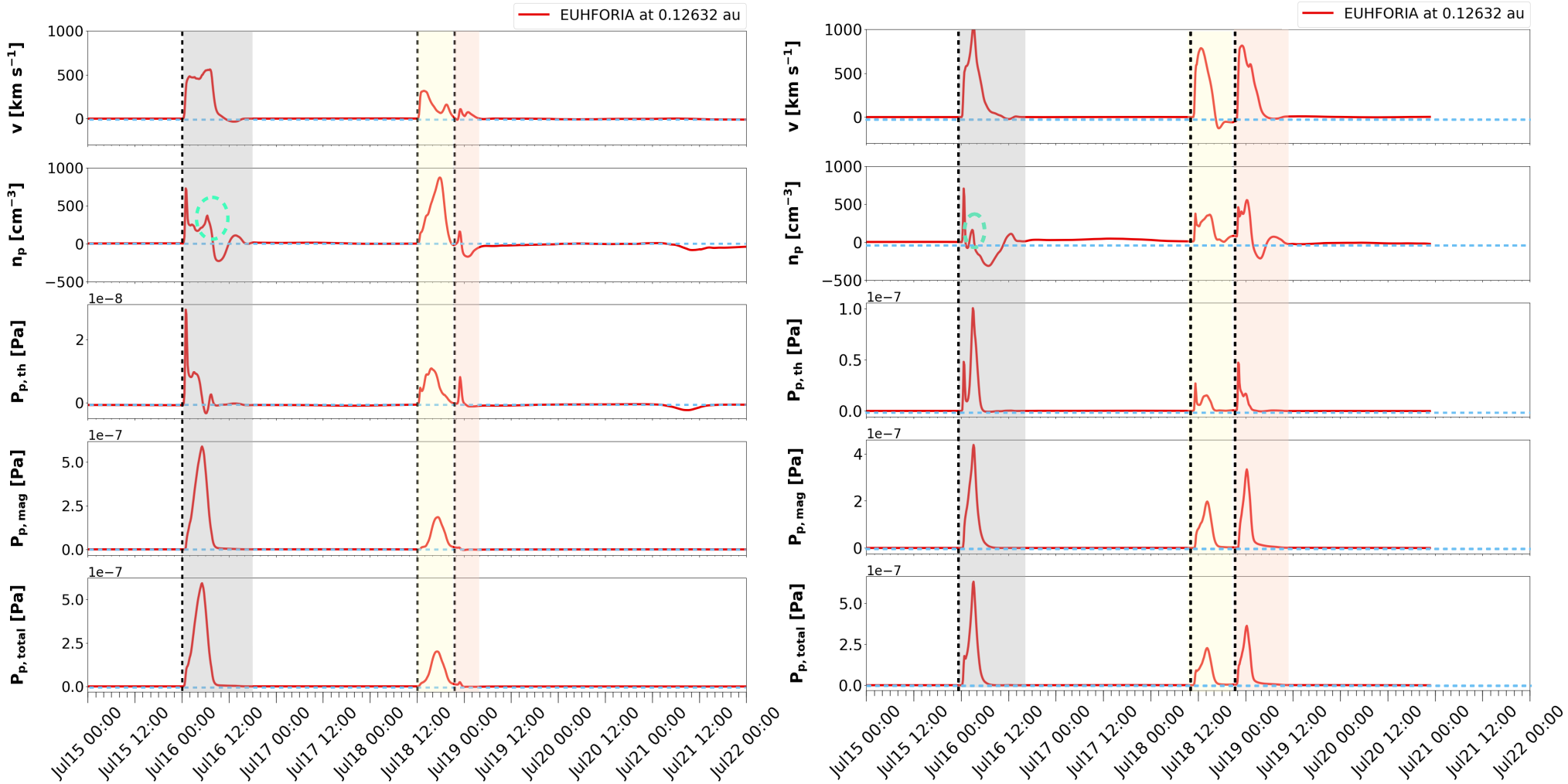}
\caption{EUHFORIA results at 0.126 au for modelling the three ICMEs with spheromaks for case 1 (left panels) and for case 2 (right panels).
From top to bottom, we show radial velocity, proton number density, and the thermal, magnetic and total pressures with the background solar wind values subtracted (see \fig{euhforia_SW}). The three ICMEs are indicated by blue, yellow and pink colors.}
 \label{fig_case1_case2_012}
\end{figure*}
\begin{figure*}[ht!]
\centering
 \includegraphics[scale=0.40]{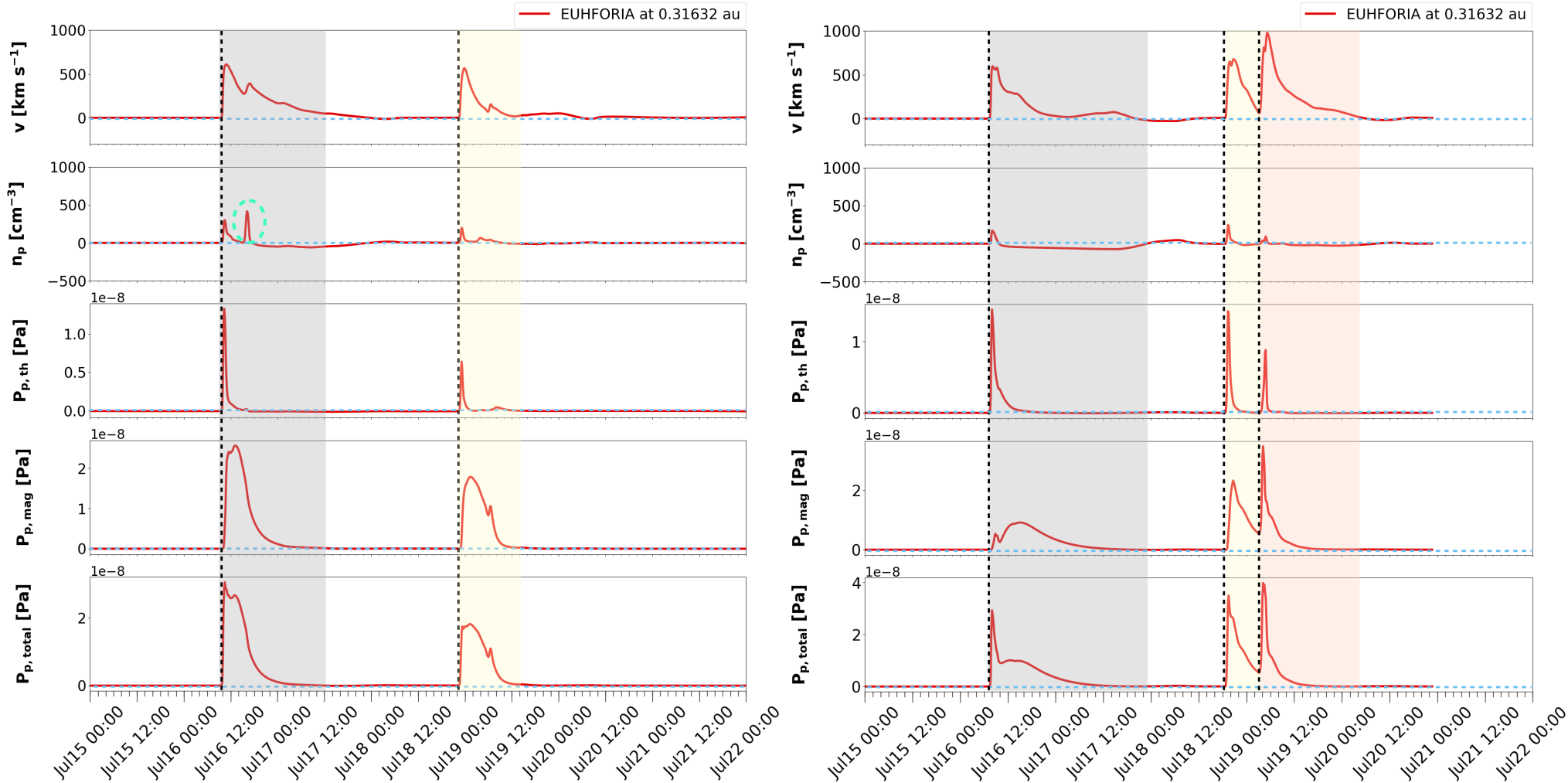}
\caption{EUHFORIA results at 0.316 au for modelling the three ICMEs with spheromaks.
The graphic convention is the same as in \fig{case1_case2_012}.
 }
 \label{fig_case1_case2_031}
\end{figure*}

\begin{figure*}[ht!]
\centering
 \includegraphics[scale=0.40]{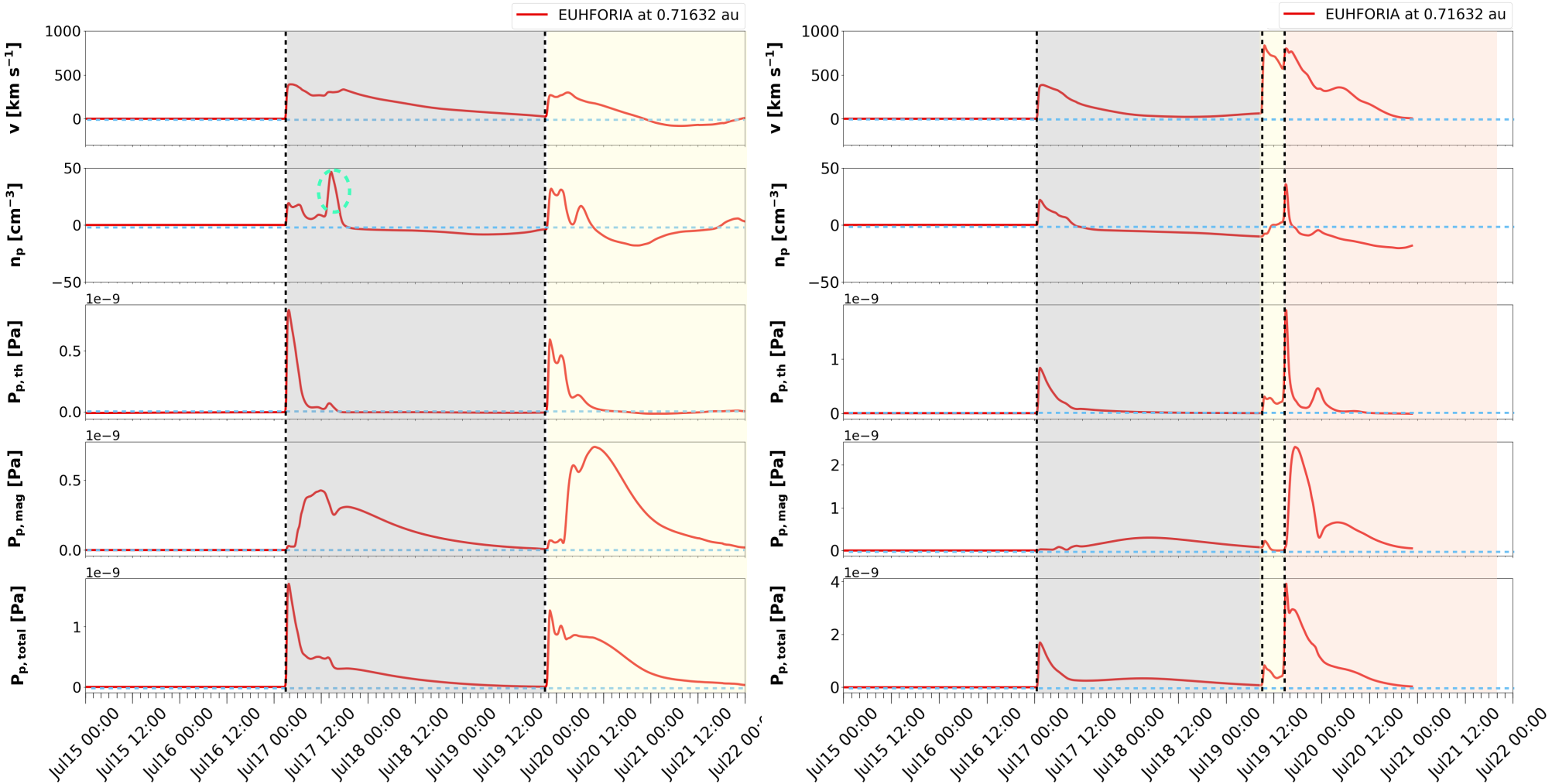}
 \caption{EUHFORIA results at 0.716 au for modelling the three ICMEs with spheromaks.
The graphic convention is the same as in \fig{case1_case2_012}.
 }
 \label{fig_case1_case2_071}
\end{figure*}

\begin{figure*}[ht!]
\centering
 \includegraphics[scale=0.40]{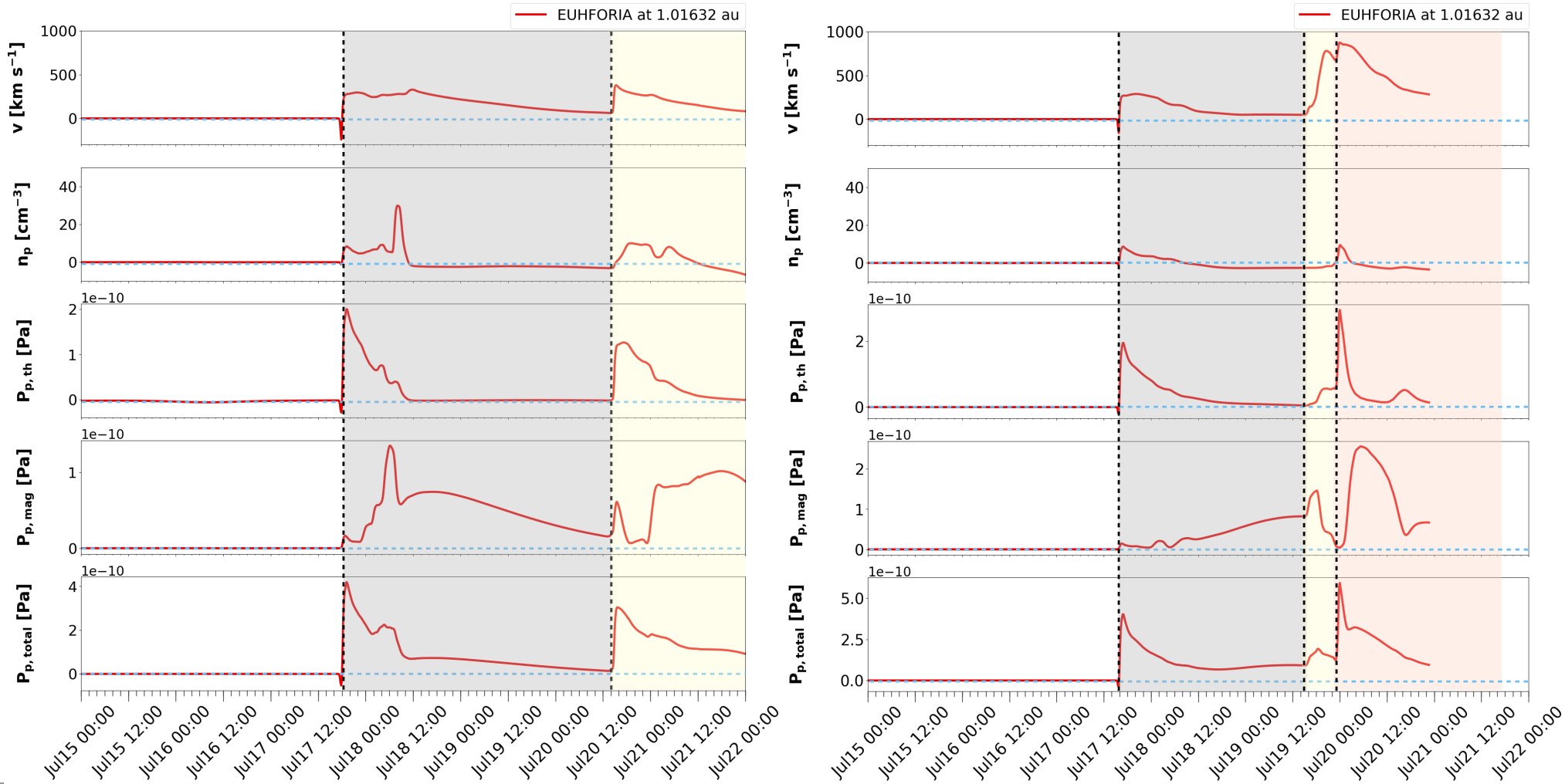}
\caption{EUHFORIA results at 1 au for modelling the three ICMEs with spheromaks.
The graphic convention is the same as in \fig{case1_case2_012}.
 }
 \label{fig_case1_case2_1}
\end{figure*}

We follow the evolution of the three ICMEs\simu\ with virtual spacecraft positioned between 0.1 and 1~au in Figures \ref{fig_case1_case2_012} through \ref{fig_case1_case2_1}. Left and right side show results for case 1 and 2, respectively. We outline the ICMEs\simu\ with different colors: grey, yellow and pink for ICME1-2-3, respectively. Note that for all figures the background solar wind has been subtracted.

The analysis of the expansion of ICME1\simuone\ at 0.126~au (Figure \ref{fig_case1_case2_012} left side), shows that the spheromak keeps a nearly uniform speed (i.e., no expansion, as set initially) before having an expansion profile (well developed after 0.2~au). This shows that the expansion is initiated between 0.126~au and 0.2~au, being a consequence of the extra total pressure inside the spheromak \textbf{with} respect to the surrounding ambient \textbf{solar wind}. A similar evolution is globally seen for ICME2\simuone\ with an expansion starting earlier on in the velocity profile. In case 2, all the ICMEs\simutwo\ get an expansion profile even earlier on. ICME1\simu\ has an initial toroidal flux stronger by a factor 1.25 in case 2 than in case 1 and an extension smaller by a factor 1.5 (see Table~\ref{tabl_icme}). This implies a magnetic field stronger by a factor 2.8. For ICME2\simu\ and ICME3\simu\ only the initial toroidal flux is stronger by a factor 1.7 in case 2 than in case 1, implying the same factor on the magnetic field strength. Then, all the ICMEs simulated in case 2 have an initial stronger internal magnetic pressure. This induces a faster expansion in case 2 than in case 1, while more moderate for ICME2-3\simu .       

A plasma density peak is formed early on at the front of all the simulated ICMEs, as expected with the compression induced by their much faster velocity than the preceding solar wind. This density peak is a sheath in formation.
Another peak is also formed with more compression for ICME1\simuone\ (dashed green ellipses in Figures \ref{fig_case1_case2_012} through \ref{fig_case1_case2_071}). In this last case the density peak survive to 1 au (see \fig{case1_case2_1}). This density peak can be interpreted as the result of the initial spheromak model that is too much out of equilibrium with its surrounding background solar wind, so that a compression is induced. Such density peak is not present in observations. Next, in the ME1, the  expansion induced a deficit of density. Also behind the ME1, the plasma is depleted because of the fast ME1 in front is inducing a large extension of this region.

During the outward evolution, the B profile becomes more and more asymmetric with a strong compression at the front and a long tail developing at the rear, in agreement with the above description of the velocity 
profiles.  The asymmetry profile result mostly from the difference of total pressure built in the ME front compared to its rear.
ICME1\simu\ has an initial B strength stronger by a factor 2.8 in case 2 than in case 1. 
This difference is no longer present at 0.126 au (see Figure \fig{case1_case2_012}). Indeed, the much stronger magnetic pressure unbalance in case 2 induces a faster ME expansion in order to reach a nearly balance of pressure with the surrounding background solar wind.  Consequently,  both the velocity and density profiles are also significantly modified at 0.126 au for case 2.

Finally, at 0.7~au the interaction between ICME1\simuone\ and ICME2\simuone\ occurs in case 1. Even at 1 au this interaction affects only the rear of ICME1, and later than observed in situ (see \fig{Case1_case2_1}). The interaction is stronger in case 2, with a larger fraction of ICME1 rear affected, in better agreement with observations. The interaction between ICME2 and ICME3 is the strongest (case 2) as it started already at about 0.16 au along the Sun--Earth direction (see associated movie). ICME2 gets fully compressed, becoming like a sheath in front of ICME3.

\begin{table}[t!]
     \centering
     
     \begin{tabular}{ccccc}
     \hline
     Case  & Distance    &  $B$ strength & $P_{\rm th}$ &  $P_{\rm mag}$\\
    & [au]    &  [nT] & [10$^{-8}$~Pa] &   [10$^{-8}$~Pa]\\
      \hline
      case 1 & 0.126 & 1000 & 3    &  60    \\
             & 0.316 &  200 & 1    &   2.5  \\
             & 0.716 &   30 & 0.08 &   0.04 \\
             & 1.016 &   10 & 0.02 &   0.01 \\
             \hline
      case 2 & 0.126 & 1000 & 10    & 50   \\  
             & 0.316 &  150 &  1    &  1   \\
             & 0.716 &   25 &  0.1  & 0.05 \\
             & 1.016 &   10 &  0.02 & 0.01 \\
       \hline
      \end{tabular}
\caption{Characteristic peak values of the magnetic field strength B, the thermal plasma pressure $P_{\rm th}$, the magnetic pressure $P_{\rm mag}$ of ICME1 at different distances from the Sun for the two studied cases.}
     \label{tabl_pressure}
\end{table}

\subsection{\bf Discussion of the pressure balance} \label{sect_evol_Discussion}

In order to understand how the ICMEs\simu\ can expand, we compute the thermal plasma pressure $\Pth$, the magnetic pressure $\Pmag$ and the total pressure $P_{\rm tot}$ of the 3 ICMEs\simu\ from 0.126~au to 1~au.  The evolution of the pressures are presented  for four distances to the Sun (see Figures \ref{fig_case1_case2_012}-\ref{fig_case1_case2_1}). The maxima of plasma and magnetic pressure peaks for ICME1\simu\ \textbf{are} presented for different distances in Table \ref{tabl_pressure}.


Already at 0.126~au the plasma pressure starts to be important in ICME1\simu\ front as the consequence of the formation of a front shock. There is not yet a significant incidence on the total pressure. However, at the next presented distance of 0.316~au, the plasma pressure starts to contribute to the total pressure in the sheath. Such contribution increases with distance, becoming the dominant pressure source at 0.716 au in the sheath of ICME1\simu.

The magnetic pressure in ME1\simu\ is much larger than the plasma pressure at 0.126 and 0.316~au.  It is remarkable that the magnetic pressure in ME1\simu\ is a factor 2.5 stronger in case 1 than in case 2 at 0.316~au, while this factor was about 1/8 initially (at 0.1 au the toroidal flux of case 2 was stronger by a factor 1.25 than this of case 1 and the magnetic flux  B$_{case2}$ = 2.8 B$_{case1}$, see section 4.3).  The much stronger initial magnetic pressure (B$^2$/2$mu_0$) in case 2 induces a much faster expansion, which implies a much weaker magnetic field closer to the Sun. This inversion of the field strength is still present at larger distances as well.

The evolution of the pressures show that the ICMEs\simu\ never {gets} in near balance pressure with their surroundings.   It is so because the front shock compression induces a larger plasma and magnetic pressure, while at the rear, the void in the background solar wind created by the fast ICME passage, implies a low total pressure.  This implies that ICME1\simu\ expands faster at 1 au ($\zeta_{simu} \approx 1.05$ for both cases) than expected with pressure balance with the surroundings ($\zeta \approx 0.73$, see Section \ref{sect_evolution}).  Next, despite having a larger initial magnetic pressure in ME1\simu\ for case 2 compared to case 1, which induces a faster expansion in the close inner heliosphere, this difference in expansion rate is erased by the time the ME1 reaches 1 au. However, the magnetic field strength is in the reverse order than it was initially (so keeping the memory of a faster expansion in the inner heliosphere).

At 1 au, the expansion of ME1\simu\ is about isotropic in both cases with a volume scaling approximately as $D^3$, compared to the volume scaling as about $D^2$ in the surrounding solar wind.  This differential expansion rate, which was even larger in the inner heliosphere, induces much lower plasma density in ME1\simu , as observed in ME1\obs , than in the solar wind. We conclude that the formation of the low density region in ME1\simu\ for the two cases is realised by setting an over magnetic pressure at 0.1~au.

In case 1, along the Sun--Earth line, ME2\simuone\ is not overtaken by ICME3\simuone\ so its evolution is roughly comparable to the one of ME1\simuone\ (up to the change in the initial parameters and the presence of ICME1\simuone\ in front which changes the encountered medium).  In case 2, ME2\simutwo\ is strongly overtaken by ICME3\simutwo\ so that its expansion, still visible in the velocity profile at 0.716 au, turns to a strong compression at 1 au. Then, the lower plasma observed in ME2\obs\ could not be reproduced in the simulation. Despite the too strong compression by ICME3\simu , even a weaker interaction is not expected to allow the presence of a so low plasma density in ME2\simu\ than observed in ME2\obs\  (lower than in ME1\obs , \fig{grison1}). 

{The previous analysis involves only insitu scalar parameters.   The precise 3D organisation of the vector magnetic in the ejecta is expected to partly affects those results.  More precisely, how they would be modified by changing the initial orientation of the spheromak axis, or by changing the initial geometry from a spheromak to a flux rope, is left for a future investigation. \citet{Asvestari2022} has made a first effort towards this goal by studying the effect of tilting of the LFF Spheromak model on an idealized background solar wind. Still, present results for scalar parameters are expected to be mostly kept in view of the results of \citet{Taubenschuss2010}, where the change of the helicity sign, and as such changing the way the flux rope interacts with the encountered field, implies roughly comparable results.}

\section{Discussion and conclusions}
\label{sect_Discussion}

In this paper, we considered a case-study of three ICMEs presenting low, and even very low, plasma density in their MEs. To model the arrival times of the three front side ICMEs occurring on 17-19 July 2002 we use the MHD EUHFORIA simulation as a tool to follow the propagation of the CMEs in the solar wind and through the inner heliosphere. We consider two cases for the spheromak boundary conditions at 0.1 au: case 1 is deduced from the original parameters in the inner corona as much as possible, and case 2 has adapted parameters to get the simulated velocity ($v$) and proton number density ($n$)  profiles closer to the ones measured at L1. 

In case 2, the increase of the initial toroidal flux within the launched spheromak by a factor 1.25 and its lateral extension by a factor 0.7 compared to case 1, induces a larger magnetic pressure, by a factor about 8,  and hence a larger expansion which accelerates the ICME front and led to a good fit for the arrival time. An enhanced initial toroidal flux, by a factor \p{1.7}, is also present initially for ICME2 and ICME3 so that they interact earlier with CME1. This gives simulated \insitu\ profiles closer to the observations at 1 au. This confirms the conclusion of the analysis of 42 CMEs measured in the inner heliosphere by two spacecraft showing that the expansion depends on the initial magnetic field strength inside the ICME and less on the magnetic field measured at L1 \citep{Lugaz2020}.   

The main difficulty generated by the original parameters (case 1) is that they do not lead to the observed arrival times.  
Moreover, we have not enough coronal data to define these parameters with a reasonable uncertainty. In particular, the initial velocity involves a crude approximation or an empirical formula derived from a limited number of CMEs \citep{Schwenn2005}, or a blend of the two. Furthermore, for the coronal observations only one viewpoint was available. Finally, the combination of an observed arrival time that is much earlier than simulated, while at the same time, a much lower arrival velocity compared to simulated, is atypical as generally late arrivals denote lower peak velocities.


We conclude that, like in the case of \citet{Scolini2021}, the injected spheromaks need to be set initially well in pressure unbalance with the surrounding solar wind in order to better reproduce the observations at 1 au. This creates a peculiar evolution in the inner heliosphere. 
Even with this abnormal initial conditions, the MHD equations bring the evolution close to what is expected at larger distances, in particular at 1 au. 
We conclude that the increase of the magnetic flux mitigates the EUHFORIA  assumptions related to a constant initial velocity of the spheromaks.

Finally, the low plasma density, present in ICME1 is well explained by an over-expansion occurring mostly in the inner heliosphere, while still partly present at 1 au. The same process is present in ICME3, while to a more moderate level.
The observed ICME2 has a very reduced plasma density (by one order of magnitude).  Since it is overtaken and compressed by ICME3, the even lower density of ICME2, than in ICME1, could not be explained by expansion. Rather, the most plausible possibility is that CME2 was coming from a coronal cavity with already low plasma density.



{\bf Acknowledgments} \\
The authors thank Dr.\ Emmanuel Chan\'e for many interesting discussions and his valuable input on this paper.  We thank Dr. Bocchialini for providing useful informations on the halo and partial coronal mass ejections.
C.V. is funded by the Research Foundation – Flanders, FWO SB PhD fellowship 11ZZ216N. BG acknowledges support of GACR Grant 22-10775S. SD acknowledges support from the argentine grants PICT-2019-02754 (FONCyT-ANPCyT) and UBACyT-20020190100247BA (UBA). E.S. acknowledges the PhD grant awarded by the Royal Observatory of Belgium. C.S. acknowledges the NASA Living With a Star Jack Eddy Postdoctoral Fellowship Program, administered by UCAR's Cooperative Programs for the Advancement of Earth System Science (CPAESS) under award no. NNX16AK22G. 
These results were also obtained in the framework of the projects C14/19/089  (C1 project Internal Funds KU Leuven), G.0D07.19N (FWO-Vlaanderen), SIDC Data Exploitation (ESA Prodex-12). EUHFORIA is developed as a joint effort between the University of Helsinki and KU Leuven. The full validation of solar wind and CME modelling is being performed within the BRAIN-be CCSOM project (Constraining CMEs and Shocks by Observations and Modeling throughout the inner heliosphere; http://www.sidc.be/ccsom/) and BRAIN-be SWiM project (Solar WInd Modeling with EUHFORIA for the new heliospheric missions). The simulations were carried out at the VSC – Flemish Supercomputer Centre, funded by the Hercules foundation and the Flemish Government – Department EWI.
This project (EUHFORIA 2.0) has received funding from the European Union’s Horizon 2020 research and innovation programme under grant agreement No 870405.
Figure 3 data analysis was performed with the AMDA science analysis system provided by the Centre de Données de la Physique des Plasmas (CDPP) supported by CNRS, CNES, Observatoire de Paris and Université Paul Sabatier, Toulouse \citep{Genot2021}.
We recognize the collaborative and open nature of knowledge creation and dissemination, under the control of the academic communit.
\\

\bibliographystyle{model5-names}
\biboptions{authoryear}
\bibliography{refs}

\end{document}